\documentclass[journal]{IEEEtran}
\usepackage{etoolbox}
\usepackage{dtsc-creafig}
\usepackage{notation}

\usepackage{epsfig}
\usepackage{color}
\usepackage{amsmath}
\usepackage{amssymb}
\usepackage{mathabx}
\usepackage{enumerate}
\usepackage{multirow}
\usepackage{bbm}
\usepackage{algorithm}
\usepackage{algorithmic}
\usepackage{lipsum,amsmath,multicol}
\usepackage{stfloats}
\usepackage{arydshln} 
\usepackage{amsfonts}
\usepackage{dsfont}
\usepackage{upgreek}
\usepackage{pgfplots} 
\usepackage{caption}
\usepackage{subfigure} % subfiguras
\usepackage{booktabs} % Top and bottom rules for tables

\definecolor{auburn}{rgb}{0.43, 0.21, 0.1}

\usepackage{cite}

\usepackage{acronym}
\acrodef{BER}{bit error rate}
\acrodef{LMMSE}{linear minimum-mean squared-error}
\acrodef{FIR}{finite impulse response}
\acrodef{CP}{cyclic prefix}

\ifCLASSINFOpdf
\else
\fi
\begin{document}
%
% paper title
% Titles are generally capitalized except for words such as a, an, and, as,
% at, but, by, for, in, nor, of, on, or, the, to and up, which are usually
% not capitalized unless they are the first or last word of the title.
% Linebreaks \\ can be used within to get better formatting as desired.
% Do not put math or special symbols in the title.
%\title{EP-based filter for turbo equalization}
%\title{{Expectation propagation for filter type turbo equalization}}
%\title{{EP for filter type turbo equalization with quadratic complexity in the filter length}}
%\title{{Expectation Propagation Turbo Equalization: a Filter-Type Implementation}}
\title{{Turbo EP-based Equalization: a Filter-Type Implementation}}
%
%
% author names and IEEE memberships
% note positions of commas and nonbreaking spaces ( ~ ) LaTeX will not break
% a structure at a ~ so this keeps an author's name from being broken across
% two lines.
% use \thanks{} to gain access to the first footnote area
% a separate \thanks must be used for each paragraph as LaTeX2e's \thanks
% was not built to handle multiple paragraphs
%

\author{Irene~Santos,
        Juan~Jos\'e~Murillo-Fuentes, Eva~Arias-de-Reyna, 
        and~Pablo~M.~Olmos% <-this % stops a space
\thanks{I. Santos, J.J. Murillo-Fuentes and E. Arias-de-Reyna are with the Dept. Teor\'ia de la Se\~nal y Comunicaciones, Escuela T. Superior de Ingenier\'i­a, Universidad de Sevilla, Camino de los Descubrimiento s/n, 41092 Sevilla, Spain. E-mail: {\tt \{irenesantos,murillo,earias\}@us.es}}% <-this % stops a space
\thanks{P. M. Olmos is with the Dept. Teor\'ia de la Se\~nal y Comunicaciones, Universidad Carlos III de Madrid, Avda. de la Universidad 30, 28911, Legan\'es (Madrid), Spain. He is also with the Instituto de Investigaci\'on Sanitaria Gregorio Mara\~n\'on (IiSGM). E-mail: {\tt olmos@tsc.uc3m.es}.}% <-this % stops a space
\thanks{This manuscript has been submitted to IEEE Transactions on Communications on September 7, 2017; revised on January 10, 2018 and March 27, 2018; accepted on April 25, 2018.  This work was partially funded by Spanish government (Ministerio de Econom\'ia y Competitividad TEC2016-78434-C3-\{2-3\}-R and Juan de la Cierva Grant No. IJCI-2014-19150) and by the European Union (FEDER).}}

% note the % following the last \IEEEmembership and also \thanks - 
% these prevent an unwanted space from occurring between the last author name
% and the end of the author line. i.e., if you had this:
% 
% \author{....lastname \thanks{...} \thanks{...} }
%                     ^------------^------------^----Do not want these spaces!
%
% a space would be appended to the last name and could cause every name on that
% line to be shifted left slightly. This is one of those "LaTeX things". For
% instance, "\textbf{A} \textbf{B}" will typeset as "A B" not "AB". To get
% "AB" then you have to do: "\textbf{A}\textbf{B}"
% \thanks is no different in this regard, so shield the last } of each \thanks
% that ends a line with a % and do not let a space in before the next \thanks.
% Spaces after \IEEEmembership other than the last one are OK (and needed) as
% you are supposed to have spaces between the names. For what it is worth,
% this is a minor point as most people would not even notice if the said evil
% space somehow managed to creep in.

% The paper headers
\markboth{ }%
{}
% The only time the second header will appear is for the odd numbered pages
% after the title page when using the twoside option.
% 
% *** Note that you probably will NOT want to include the author's ***
% *** name in the headers of peer review papers.                   ***
% You can use \ifCLASSOPTIONpeerreview for conditional compilation here if
% you desire.

% If you want to put a publisher's ID mark on the page you can do it like
% this:
%\IEEEpubid{0000--0000/00\$00.00~\copyright~2015 IEEE}
% Remember, if you use this you must call \IEEEpubidadjcol in the second
% column for its text to clear the IEEEpubid mark.

% use for special paper notices
%\IEEEspecialpapernotice{(Invited Paper)}

% make the title area
\maketitle

% As a general rule, do not put math, special symbols or citations
% in the abstract or keywords.
\begin{abstract}
%We propose a novel filter type equalizer to quite improve the solution of the \ac{LMMSE} turbo equalizer, with computational complexity constrained to be quadratic in the filter length. When transmitting over channels affected with inter-symbol interference, the optimal BCJR equalization is computationally unfeasible whether high-order modulations and/or large memory channels are used. In this scenario, the estimation using a \ac{FIR} filter under the LMMSE criterion and turbo-equalization exhibits a  behavior robust to changes in channel realizations and constellations used. In this paper, we show that this solution can be significantly improved by using expectation propagation (EP) in the estimation of the a posteriori probabilities. It yields a more accurate estimation of the extrinsic distribution to be sent to the channel decoder constraining the computational complexity to be quadratic in the length of the FIR. Simulation results included show that this new EP-based filter remarkably outperforms the LMMSE approach, with and without turbo equalization. 

%%%% 
We propose a novel filter-type equalizer to improve the solution of the \ac{LMMSE} turbo equalizer, with computational complexity constrained to be quadratic in the filter length. When high-order modulations and/or large memory channels are used the optimal BCJR equalizer is unavailable, due to its computational complexity. In this scenario, the filter-type LMMSE turbo equalization exhibits a good performance compared to other approximations. 
%
%When transmitting over channels affected with inter-symbol interference, the optimal BCJR equalization is computationally unfeasible whether high-order modulations and/or large memory channels are used. In this scenario, the estimation using a \ac{FIR} filter under the LMMSE criterion and turbo-equalization exhibits a  behavior robust to changes in channel realizations and constellations used. 
%
In this paper, we show that this solution can be significantly improved by using expectation propagation (EP) in the estimation of the a posteriori probabilities. First, it yields a more accurate estimation of the extrinsic distribution to be sent to the channel decoder. Second, compared to other solutions based on EP the computational complexity of the proposed solution is constrained to be quadratic in the length of the \ac{FIR}. 
%This reduction in the complexity is the main advantage in comparison with other EP equalizers, such as the batch or block-EP approach. 
In addition, we review previous EP-based turbo equalization implementations. Instead of considering default uniform priors we exploit the outputs of the decoder. %, with improved performance.
%by exploiting the output of the decoder to determine the priors used at  each execution of the EP algorithm, instead of considering default uniform priors. 
%, once the turbo procedure starts.
%by considering that the priors computed from the decoder are not uniforms any more once the turbo procedure starts. 
%We also study the EP parameters to obtain a reduced-complexity version for the turbo approach of the algorithm which is three times lower in terms of complexity than previous approaches. 
Some simulation results are included to show that this new EP-based filter remarkably outperforms the turbo approach of previous versions of the EP algorithm and also improves the LMMSE solution, with and without turbo equalization.

%We investigate an alternative exchange of information between equalizer/decoder when performing turbo-equalization. We propose to use a posteriori probabilities instead of extrinsic ones. 
%A linear minimum mean squared error (LMMSE) equalizer is compared with an expectation propagation (EP) approach in both situations. Simulation results show their performance for different scenarios. 
%After experimental comparison of the standard turbo-equalization with extrinsic and the turbo-equalization with posteriors, we found that there are no improvements of one compared to the other for large constellations. But we do get an improvement with extrinsic information in BPSK.
\end{abstract}

% Note that keywords are not normally used for peerreview papers.

% For peer review papers, you can put extra information on the cover
% page as needed:
% \ifCLASSOPTIONpeerreview
% \begin{center} \bfseries EDICS Category: 3-BBND \end{center}
% \fi
%
% For peerreview papers, this IEEEtran command inserts a page break and
% creates the second title. It will be ignored for other modes.
\IEEEpeerreviewmaketitle

\section{Introduction}
% The very first letter is a 2 line initial drop letter followed
% by the rest of the first word in caps.
% 
% form to use if the first word consists of a single letter:
% \IEEEPARstart{A}{demo} file is ....
% 
% form to use if you need the single drop letter followed by
% normal text (unknown if ever used by the IEEE):
% \IEEEPARstart{A}{}demo file is ....
% 
% Some journals put the first two words in caps:
% \IEEEPARstart{T}{his demo} file is ....
% 
% Here we have the typical use of a "T" for an initial drop letter
% and "HIS" in caps to complete the first word.
\IEEEPARstart{M}{any} digital communication systems need to transmit over channels that are affected by inter-symbol interference (ISI). 
%\notaI{This effect can be mitigated by using a time or frequency domain equalizers. The latter ones, such as OFDM, delete the ISI by introducing  a guard interval between OFDM symbols by means of a ciclic prefix (CP), whose length must be at least the one of the channel. However, for large channel dispersions and short frame lengths, it yields a lost in band width. In addition, the ISI problem is replaced by intercarrier interference (ICI) due to the time variation \cite{Tuchler11}. For these reasons, we focus on the time domain equalizers. }
%This effect can be mitigated by using an equalizer. 
The equalizer produces a probabilistic estimation of the transmitted data given the vector of observations \cite{Haykin09}. Significant improvements are found when the previous estimation is given to a probabilistic channel decoder \cite{Salamanca12}. \notajj{Equalization can be done in the frequency domain to avoid complexity problems associated with the inverse of covariance matrices \cite{Karjalainen07}.} In addition, feeding the equalizer back again with the output of the decoder, iteratively, yields a turbo-equalization scheme that significantly reduces the overall error rate \cite{Douillard95,Koetter02,Koetter04}. 

The BCJR algorithm \cite{Bahl74} performs optimal turbo equalization under the maximum a posteriori (MAP) criterion. It provides a posteriori probability (APP) estimations given some \notaI{\textit{a priori}} information about the transmitted data. However, its complexity grows exponentially with the length of the channel and the constellation size, becoming intractable for few taps and/or multilevel constellations. 
In this situation, approximated BCJR solutions, such as \cite{franz98,Colavolpe01,Sikora05,Fertonani07}, can be used. They are based on a search over a simplified trellis with only $\Me$ states, yielding a complexity which is linear in this number of states. However, the performance of these approaches 
is quite dependent on the channel realization and the order of the constellation used. {In addition, these approximated BCJR solutions degrade rapidly} if the number of survivor paths does not grow according to the total number of states. For these reasons, filter-based equalizers are preferred \cite{Berrou10}.

 {A quite extended} filter {type} equalizer in the literature is {based on} the well-known linear minimum-mean squared-error (LMMSE) algorithm \cite{Singer01,Koetter02,Tuchler11}. 
{This LMMSE filter is an appealing alternative where the BCJR is computationally unfeasible due to its robust performance with} linear complexity in the frame length, $\tamframe$, and quadratic dependence with the window length, $\winsize$. {From a Bayesian point of view, the LMMSE algorithm obtains a Gaussian extrinsic distribution by replacing the discrete prior distribution of the transmitted symbols with a Gaussian prior. }

{A more accurate estimation for the extrinsic distribution can be obtained by replacing the prior distributions with approximations of the probability distribution. This can be done by means of the expectation propagation (EP) algorithm. The EP approach projects the approximated posterior distribution into the family of Gaussians by matching its moments iteratively with the ones of the true posterior.  
This algorithm has been already successfully applied to \notajj{multiple-input multiple-output (MIMO)} systems \cite{Cespedes14-2} and low-density parity-check (LDPC) channel decoding \cite{Olmos12,Salamanca13}{, among others}. {It has been also applied to turbo equalization in a message passing approach as a way to incorporate into the BP algorithm the discrete information coming from the channel decoder \cite{Hu06,Sun15}. {These message passing methods reduce to the LMMSE estimation if no turbo equalization is employed}. A different approach is proposed in \cite{Santos15,Santos16} {under the name of block EP (BEP) } where, rather than applying EP after the channel decoder, it is used {within the equalizer to better approximate the posterior, outperforming previous solutions.} 
%LMMSE performance even without turbo equalization. 

The computational complexity of previous EP-based equalizers is large for long frame lengths or memories of the channel. Due to its block implementation, the complexity of the BEP} is quadratic in the frame length, becoming intractable for large frames \cite{Santos16}. To overcome this drawback, a smoothing EP (SEP) implementation is proposed in \cite{Santos17}, but its complexity is cubic with the memory of the channel. Furthermore, due to their iterative procedure, their computational load is roughly $\iterep$ times the one of the LMMSE counterparts, where $\iterep$ is the number of iterations used in the EP algorithm, typically around 10 {\cite{Cespedes14-2,Santos15,Santos16}}. } Besides, in both, BEP and SEP, uniform discrete priors are assumed for the constellation of the modulations when computing the EP approximations, even within the turbo equalization iterations, while the use of information from the decoder remains unexplored. 

%. We investigate if information from the decoder could be used as priors in these projections, outperforming the previous EP approaches.  
%However, once the turbo procedure starts, not allowing the algorithms to get the true moments of the distribution. Previous solutions considered uniform priors in the EP projections even in the turbo equalization iterations. The turbo equalization procedure with EP can be improved by   

%{It has been also applied to simple and turbo equalization in a block approach (BEP) \cite{Santos15,Santos16}, where BEP exhibits a quite robust performance, regardless of the channel response and the constellation used, and quite outperforms the LMMSE approach with higher gains.}  However, due to its block implementation, the complexity is quadratic in the frame length, becoming intractable for large frames. 
%A different implementation is proposed in \cite{Hu06,Sun15}, where the EP algorithm is used in a message passing approach as a way to incorporate into the BP algorithm the discrete information coming from the channel decoder. In this paper, rather than applying EP after the channel decoder, we use it before the channel decoder, including the discrete nature of symbols even before the turbo procedure.  {Si nos publican en el ICC, debemos citar el paper aqui. }

The results developed in this paper focus on improving these previous EP-based equalizers \cite{Santos16,Santos17} both in computational complexity and performance. 
%
%{We apply the EP algorithm in a filter-based approach, keeping some of the advantages of both block and filtered approaches \cite{Santos16,Santos17} and improving their turbo approach. 
% We investigate if information from the decoder could be used as priors in these projections, outperforming the previous EP approaches.  
\notaI{First, we improve the prior information used in the equalizer once the turbo procedure starts, forcing the true discrete prior to be non-uniform in contrast to the uniform priors used by previous EP-based approaches.}
%First, we investigate if, along the turbo equalization iterations, the information from the decoder could be used as non-uniform discrete priors in the EP approximations. 
As a result, we achieve a performance improvement. Second, the computational complexity of the EP algorithm is reduced to roughly a third part of that in \cite{Santos16}, by optimizing the choice of EP parameters. Third, and most important, a {new}  filter-type EP solution is designed. This solution is constrained to have linear complexity in the frame length and quadratic in the filter length, i.e., {it is endowed with} the same complexity order than the LMMSE filter.  

The {novel} EP-based filter {proposed} outperforms the LMMSE algorithm with a robust behavior {to changes in the constellation size and the channel realization}, as the BEP and SEP approaches do \cite{Santos16,Santos17}. 
{In the experiments included, we show that the EP filter solution greatly improves the LMMSE solution with and without turbo equalization, specifically we have 2 dB gains for a BPSK, 5 dB for the 8-PSK and 6-13 dB for 16 and 64-QAM, respectively. In comparison with previous EP approaches, the EP filter matches their performance with BPSK constellations, and outperforms them with gains of 2 dBs for 8-PSK and 4-5 dBs for 16 and 64-QAM. We study the extrinsic information transfer (EXIT) charts 
\noteE{\cite{Koetter02,Kansanen07}} of our proposal for a BPSK, where the EP-based filter achieves the same performance as the BEP}. 

The scope of this paper encompasses time domain equalization. Frequency domain equalization has received a lot of attention as it usually achieves a complexity reduction \notaIc{for the block-wise processing }\cite{Falconer02,Tuchler00,Tuchler11,Karjalainen07}. For this reason, derivation of a frequency domain counterpart for the proposed EP based turbo-equalizer remains as a future research line. \notaIc{Another promising research route is the application to MIMO with channels with memory \cite{Veselinovic05, Karjalainen07,Xiao17}. }

The paper is organized as follows. We first describe in \SEC{sysmod} the model of the communication system at hand. \SEC{EP} is devoted to develop a new implementation of the EP-based equalizer considering non-uniform priors and studies the optimal values for the parameters. In \SEC{Filter}, we review the formulation for the LMMSE filter in turbo equalization and describe the novel EP filter-type solution proposed. 
In \SEC{sim}, we include several \notaIc{simulations} to compare both EP and LMMSE approaches. {We end with conclusions.}

{Through the paper, %We use the following specific notation throughout the paper. We 
we denote} the $i$-th entry of a vector $\vect{\beforechannel}$ as 
$\beforechannel_i$, \notaI{its complex conjugate as $\vect{\beforechannel}\cnj$ and its Hermitian transpose as $\matr{u}\her$}. \notaI{We define $\delta(u_i)$ as the delta function that takes value one if $u_i=0$ and zero in other case. }We use $\cgauss{\vect{\beforechannel}}{\boldsymbol{\upmu}}{\boldsymbol{\Sigma}}$ to denote a normal distribution of a random \notaI{proper complex} vector %$\beforechannel$ with mean $\mu$ and variance $\sigma^2$ we use the notation $\gauss{\beforechannel}{\mu}{\sigma^2}$. In case of a random vector 
$\vect{\beforechannel}$ with mean vector $\boldsymbol{\upmu}$ and covariance  matrix $\boldsymbol{\Sigma}$.  %The expression $\mbox{Proj}_G[q(\cdot)]=\gauss{\cdot}{m}{\sigma^2}$ is the projection of the distribution given as an argument, $q(\cdot)$, with moments $m$ and $\sigma^2$, respectively, into the family of Gaussians \cite{Sun15}.  

\section{System Model}\LABSEC{sysmod}

\begin{figure*}[htb]
\centering
\includegraphics[width=7.0625in]{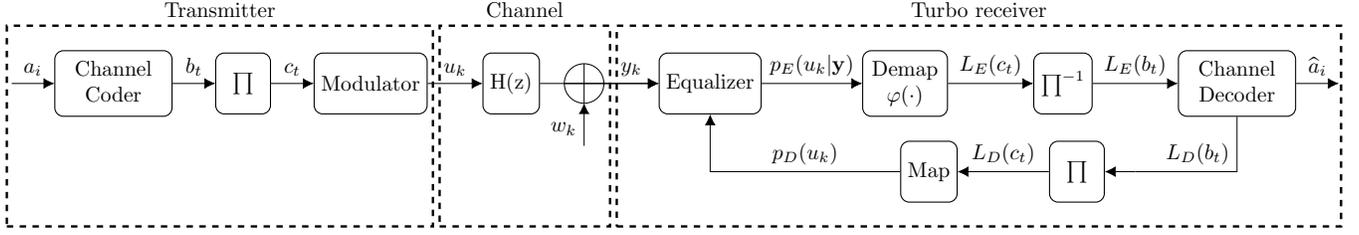}
%\scalebox{0.84}{\input{pics/SysModelTurbo3.tex}}
\caption{\small System model.} \LABFIG{sysmod}
\end{figure*}

{The model of the communication system is depicted in \FIG{sysmod}, including turbo equalization at the receiver}. There are three main blocks: transmitter, channel and turbo receiver. 
%Without loss of generality, in the following we adopt the real-valued channel model with PAM constellations. Similar results can be readily formulated for the complex case with doubled dimension where real and imaginary parts are stacked.

\subsection{Transmitter}

The information bit sequence, \notaI{$\vect{a}=[\allvect{a}{1}{\k}]\trs$} where $\notaI{a_i}\in\{0,1\}$, is encoded into the coded bit vector $\vect{b}=[\allvect{b}{1}{\t}]\trs$ with a code rate equal to $\rate=\k/\t$. After permuting the bits with an interleaver, the codeword $\vect{c}=[\allvect{c}{1}{\t}]\trs$ {is partitioned into $\tamframe$ blocks of length $Q=\log_2(\modsize)$, $\vect{c}=[\vect{c}_{1}, ..., \vect{c}_{\tamframe}]\trs$ where $\vect{c}_{\iter}=[c_{\iter,1}, ..., c_{\iter,Q}]$,  and modulated with a \notaI{complex} $\modsize$-ary constellation $\mathcal{A}$ of size $|\mathcal{A}|=\modsize$. % to obtain a sequence of $\tamframe=\lceil \t/ \log_{2}\modsize \rceil$ symbols. %, where each symbol is a combination of $\log_2(\modsize)$ . 
These modulated symbols,  $\vect{\beforechannel}=[\allvect{\beforechannel}{1}{{\tamframe}}]\trs$, where each component  $\beforechannel_\iter=\notaI{\re(\beforechannel_\iter)+j\im(\beforechannel_\iter)} \in\mathcal{A}$, are transmitted over the channel.  
%$\beforechannel_\iter=\re(\beforechannel_\iter)+\text{j}\im(\beforechannel_\iter)\in\mathcal{A}$.  %and ${\tamframe}'=n{\tamframe}$ for some positive integer $n$. 
Hereafter, %$\mathcal{A}$ denotes the set of symbols of the constellation of order $|\mathcal{A}|=\modsize$ and the average
transmitted symbol energy and energy per bit are denoted as $\energy$ and $E_b$, respectively.}
%, hence the alphabet of $\vect{\beforechannel}$ symbols has size $|\mathcal{A}|^\tamframe$. 
%{We assume that the constellation is normalized, i.e., the mean transmitted symbol energy is equal to one. } We denote the energy per bit by $E_b$. 

\subsection{Channel}
The channel is completely specified by the CIR, i.e., $\vect{h}=[\allvect{h}{1}{\ntaps}]\trs$, where $\ntaps$ is the number of taps, and is corrupted with AWGN whose noise variance, $\sigma_\noise^2$, is known. %, that we assume known at the receiver. 
%The received signal $\vect{\afterchannel}=[\allvect{\afterchannel}{1}{{\tamframe+\ntaps-1}}]\trs$ 
%%\in\CC^{\tamframe+\ntaps-1}$
% is given by
% 
Each $\iter$-th entry of the \notaI{complex} received signal $\vect{\afterchannel}=[\allvect{\afterchannel}{1}{{\tamframe+\ntaps-1}}]\trs$  is given by
%$\matr{H}$ is a circulant matrix, % the $\iter$-th received entry is given by
\begin{equation}\LABEQ{yk}
\afterchannel_\iter=\sum_{\notaI{j}=1}^{\ntaps} h_{\notaI{j}}\beforechannel_{\iter-\notaI{j}+1} + \noise_\iter= \vect{h}\trs\vectcomp{\beforechannel}{\iter}{{\iter-\ntaps+1}}+\noise_\iter,
\end{equation}
where ${\noise_\iter}\sim\cgauss{{\noise_\iter}}{{0}}{\sigma_\noise^2}$ and $\beforechannel_\iter=0$ for $\iter<1$ and $\iter>\tamframe$.

\subsection{Turbo receiver}
\notaI{When no information is available from the channel decoder, the posterior probability of the transmitted symbol vector $\vect{\beforechannel}$ given the whole vector of observations $\vect{\afterchannel}$ yields
\begin{equation}\LABEQ{pugiveny}
p(\vect{\beforechannel}|\vect{\afterchannel})\; \propto \;{p(\vect{\afterchannel}|\vect{\beforechannel})p(\vect{\beforechannel})} 
\end{equation}
where, assuming equiprobable symbols, the prior would be given by 
\begin{equation}\LABEQ{indfunc2}
p(\vect{\beforechannel})=\frac{1}{\modsize}\prod_{\iter=1}^\tamframe\sum_{\beforechannel\in\mathcal{A}} \delta(\beforechannel_\iter-\beforechannel). 
\end{equation}
This prior matches with the definition given in \cite{Santos16} but, as explained below, it is just valid before the turbo procedure. }

\notaI{In a turbo architecture the equalizer and decoder iteratively exchange information for the same set of received symbols \cite{Koetter02,Tuchler11}. Traditionally, this exchange of information is done in terms of extrinsic probabilities in order to improve convergence and avoid instabilities. 
The extrinsic information at the output of the equalizer (see \FIG{sysmod}), $p_E(\beforechannel_\iter|\vect{\afterchannel})$, is computed so as to meet the turbo principle \cite{Singer01}. These probabilities, $p_E(\beforechannel_\iter|\vect{\afterchannel})$, are approximated when the optimal solution is intractable. We will denote the approximation by $q_E(\beforechannel_\iter)$.}

\notaI{The extrinsic distributions are demapped, 
\begin{equation}\LABEQ{LLRdemap}
L_E(c_{\iter,\iterj})=\log \frac{\sum_{\beforechannel_\iter\in\mathcal{A} | c_{\iter,\iterj}=0} p_E(\beforechannel_\iter|\vect{\afterchannel})}{\sum_{\beforechannel_\iter\in\mathcal{A} | c_{\iter,\iterj}=1} p_E(\beforechannel_\iter|\vect{\afterchannel})}, 
\end{equation}
deinterleaved and given to the decoder as extrinsic log-likelihood ratios, $L_E(b_t)$. 
The channel decoder computes an estimation of the information {bits, $\widehat{\vect{a}}$, along with the extrinsic LLRs on the coded bits}, computed as 
\begin{align}\LABEQ{extdec}
L_D(b_t|L_E(\vect{b}))=\log\frac{p(b_t=0|L_E(\vect{b}))}{p(b_t=1|L_E(\vect{b}))}{-L_E(b_t)}.
\end{align}
These extrinsic LLRs are interleaved, mapped again and given to the equalizer as updated priors, $p_D(\vect{\beforechannel}|L_E(\vect{b}))$, which are computed as 
\begin{equation}\LABEQ{indfunc}
p_D(\beforechannel_\iter|L_E(\vect{b}))=\sum_{\beforechannel\in\mathcal{A}} \delta(\beforechannel_\iter-\beforechannel) \prod_{\iterj=1}^{Q} p_D(c_{\iter,\iterj}=\varphi_\iterj(\beforechannel)|L_E(\vect{b})),
\end{equation}
with $\varphi_\iterj(\beforechannel)$ denoting the $\iterj$-th bit associated to the demapping of symbol $\beforechannel$. This process is repeated iteratively for a given maximum number of iterations, $\nturbo$, or until convergence. 
Note that in \FIG{sysmod} we have included  the computation of the extrinsic information within the equalization and channel decoding blocks. 
Note also that, once the turbo procedure starts, the prior in \EQ{pugiveny} is conditioned on the input at the channel decoder and the symbols are not equiprobable anymore. In this situation, the posterior distribution computed by the equalizer is given by
\begin{equation}\LABEQ{pugiveny2}
p(\vect{\beforechannel}|\vect{\afterchannel})\; \propto \;{p(\vect{\afterchannel}|\vect{\beforechannel}) \prod\limits_{\iter=1}^{\tamframe} p_D(\beforechannel_\iter|L_E(\vect{b}))}, 
\end{equation}
The true posterior distribution in \EQ{pugiveny} and \EQ{pugiveny2} has complexity proportional to $\modsize^\ntaps$. When this complexity becomes intractable, we will approximate it, denoting it as  $q(\vect{\beforechannel})$. 
In the following, we omit the dependence on the input at the decoder, $L_E(\vect{b})$, to keep the notation uncluttered in the rest of the paper. It is also uncluttered in \FIG{sysmod}. }

\section{Non-uniform BEP turbo equalizer}\LABSEC{EP}

%, while in this paper we rather exploit the information from the decoder to define a particular nonuniform prior for each symbol. 
%

EP \cite{Minka01thesis,Minka01,Seeger05,Bishop06,Takeuchi17} is a technique in Bayesian machine learning that approximates a (non-exponential) distribution with an exponential distribution whose moments match the true ones. In this paper, we focus on computing a Gaussian approximation for the posterior in \EQ{pugiveny2}, which is clearly non Gaussian due to the product of discrete priors in \EQ{indfunc}. 
%
%As introduced in \cite{Santos16}, the EP algorithm allows to find an accurate global Gaussian approximation, $q(\vect{\beforechannel}|\vect{\afterchannel})$, to the true posterior in \EQ{pugiveny}, $p(\vect{\beforechannel}|\vect{\afterchannel})$, which optimally matches the moments of both distributions. % by minimizing the Kullback Liebler 
%
%The only non-gaussian term in \EQ{pugiveny} is the prior 
As introduced in \cite{Santos16}, \notajj{  this is done by iteratively updating an approximation  within the Gaussian exponential family by replacing the non Gaussian prior terms in \EQ{pugiveny} by a product of Gaussians\footnote{Note that in \cite{Santos16} we used an alternative expression for  \EQ{aproxindicator} (an exponential distribution with parameters $\gamma_\iter=m_\iter/\notaI{\eta}_\iter$ and $\Lambda_\iter=1/\notaI{\eta}_\iter$).  }, i.e., 
\begin{align}\LABEQ{aproxindicator}
%\begin{equation}\LABEQ{aproxepblock}
\aproxfunc^{[\ell]}(\vect{\beforechannel}) & \;\propto\; 
p(\vect{\afterchannel}|\vect{\beforechannel}) {\prod\limits_{\iter=1}^{\tamframe} \tilde{p}_D^{[\ell]}({\beforechannel_\iter})}
\nonumber\\
&=\cgauss{\vect{\afterchannel}}{\matr{H}\vect{\beforechannel}}{\std{\noise}^{2}\matr{I}}{\prod\limits_{\iter=1}^{\tamframe} \cgauss{\beforechannel_\iter}{m_\iter^{[\ell]}}{\eta_\iter^{[\ell]} }}
%
%\cgauss{\vect{\afterchannel}}{\matr{H}\vect{\beforechannel}}{\std{\noise}^{2}\matr{I}} \prod_{\iter=1}^{\tamframe} \cgauss{\beforechannel_\iter}{m_\iter^{[\ell]}}{\notaI{\eta}_\iter^{[\ell]} },
%\end{equation}
%\mathbb{P}_{\beforechannel_\iter\in\mathcal{A}} \leftarrow 
%p(\vect{\beforechannel}) \stackrel{\mbox{approx}}{\longleftrightarrow} \prod_{\iter=1}^\tamframe \cgauss{\beforechannel_\iter}{m_\iter^{[\ell]}}{v_\iter^{[\ell]} }
%q_\iter^{[\ell]}(\beforechannel_\iter) = \cgauss{\beforechannel_\iter}{m_\iter^{[\ell]}}{v_\iter^{[\ell]} }.
\end{align}
}
%\notajj{Revisar por que sale m y v como si fueran vectores}
The marginalization of the resulting approximated Gaussian posterior distribution for the $\iter$-th transmitted symbol and $\ell$-th EP iteration yields
\begin{equation}\LABEQ{postnubep}
q^{[\ell]}(\beforechannel_\iter)\sim \cgauss{\beforechannel_\iter}{\mu_\iter^{[\ell]}}{s_\iter^{2[\ell]} }
\end{equation}
where
\begin{align}
\mu_\iter^{[\ell]}&=m_\iter^{[\ell]}+ \LABEQ{posmeanep}\\
&+\notaI{\eta}_\iter^{[\ell]}\vecti{h}{\iter}\her \left(\sigma_\noise^2\matr{I}+\matr{H}\mbox{diag}(\notaI{\boldsymbol{\eta}}^{[\ell]})\matr{H}\her\right)\inv (\vect{\afterchannel}-\matr{H}\vect{m}^{[\ell]}), \nonumber \\
s_\iter^{2[\ell]} & = \notaI{\eta}_\iter^{[\ell]}-\notaI{\eta}_\iter^{2[\ell]}\vecti{h}{\iter}\her \left(\sigma_\noise^2\matr{I}+\matr{H}\mbox{diag}(\notaI{\boldsymbol{\eta}}^{[\ell]})\matr{H}\her\right)\inv \vecti{h}{\iter},  \LABEQ{posvarnep}
\end{align}
$\matr{H}$ is the $\tamframe+\ntaps-1\times\tamframe$ channel matrix given by
\begin{align}\LABEQ{smat}
\matr{H}=
\begin{bmatrix}
h_1 & 0& \hdots & 0\\
\vdots & \ddots & \ddots &  \vdots \\
h_{\ntaps} &  & \ddots & 0\\
0 & \ddots &  & h_1\\
\vdots & \ddots& \ddots & \vdots \\
 0 & \hdots & 0 & h_\ntaps
\end{bmatrix}
\end{align}
and $\vecti{h}{\iter}$ is the $\iter$-th column of $\matr{H}$ (see \APEN{ap1} for the demonstration). 
At this point it is interesting to remark that \EQ{postnubep}-\EQ{posvarnep} are completely equivalent to equations (15)-(17) in \cite{Santos16}. \notaI{Here we developed the values of the mean and variance for each symbol while in  \cite{Santos16} they were computed in block form.} 
%
%The only difference is that, here, we obtain the values of the mean and variance independently for each symbol while in \cite{Santos16} it is done in block form. 
The current description is simpler because we only include the elements of the covariance matrix that are used during the execution of the algorithm, excluding the non-diagonal elements.
%We are not including the non-diagonal elements of the covariance matrix, because they are not used during the execution of the algorithm. 

The mean and variance parameters in \EQ{aproxindicator} are initialized with the statistics from the channel decoder as
\begin{align}\LABEQ{meanturbo}
m_\iter^{{[1]}}&=\sum_{\beforechannel\in\mathcal{A}} \beforechannel \cdot p_D(\beforechannel_\iter=\beforechannel), \\
\notaI{\eta}_\iter^{{[1]}}&=\sum_{\beforechannel\in\mathcal{A}}  (\beforechannel-m_\iter^{{[1]}})\cnj(\beforechannel-m_\iter^{{[1]}}) \cdot p_D(\beforechannel_\iter=\beforechannel). \LABEQ{varturbo}
\end{align}
Then, they are updated in parallel and iteratively by matching the moments of the following distributions
\begin{equation}\LABEQ{MM}
q_E^{[\ell]}(\beforechannel_\iter)p_D(\beforechannel_\iter) \stackrel{\mbox{\begin{tabular}{c}moment\\matching\end{tabular}}}{\longleftrightarrow} q_E^{[\ell]}(\beforechannel_\iter)\cgauss{\beforechannel_\iter}{m_\iter^{[\ell{+1}]}}{\notaI{\eta}_\iter^{[\ell{+1}]} }
\end{equation}
where $q_E^{[\ell]}(\beforechannel_\iter)$ is an extrinsic marginal distribution computed as\footnote{\notaI{Note that in this paper we used $q_E^{[\ell]}(\beforechannel_\iter)$ to denote the extrinsic marginal distribution, while in \cite{Santos16} we denoted as $q^{[\ell]\backslash \iter}(\beforechannel_\iter)$ and called it cavity marginal function. }} %\footnote{Note that in this paper we used $t_k$ to denote the variance of \EQ{extrinsicnubep}, while in \cite{Santos16} we denoted it with $v_k^2$. }
\notajj{
\begin{align}\LABEQ{extrinsicnubep}
q_E^{[\ell]}(\beforechannel_\iter)= q^{[\ell]}(\beforechannel_\iter)/\tilde{p}_D^{[\ell]}({\beforechannel_\iter})=
\cgauss{\beforechannel_\iter}{z_\iter^{{[\ell]}}}{\notaI{v_\iter^{{2[\ell]}}}}
\end{align} }
where
\begin{align}\LABEQ{mean_ext}
z_\iter^{[\ell]}&=\frac{\mu_\iter^{[\ell]}\notaI{\eta}_\iter^{[\ell]}-m_\iter^{[\ell]}s_\iter^{2[\ell]}}{\notaI{\eta}_\iter^{[\ell]}-s_\iter^{2[\ell]}}, \\
\notaI{v_\iter^{2[\ell]}}&=\frac{s_\iter^{2[\ell]}\notaI{\eta}_\iter^{[\ell]}}{\notaI{\eta}_\iter^{[\ell]}-s_\iter^{2[\ell]}} . \LABEQ{var_ext}
\end{align}
\notaI{Note that this equalizer differs from the one in \cite{Santos16} because we used different definitions for the true prior,  $p_D({\beforechannel_\iter})$. In the current manuscript, we considered non-uniform and discrete priors, given by \EQ{indfunc}, during the moment matching procedure in the equalizer, while in \cite{Santos16} uniform priors as in \EQ{indfunc2} were considered by default even after the turbo procedure. }
{To increase the accuracy of the algorithm, a damping procedure follows the moment matching in \EQ{MM}. We have defined an algorithm, described in \ALG{MMD}, called {\it{Moment Matching and Damping}} that runs these two procedures. }

\begin{algorithm}[!tb]
\begin{algorithmic}
\STATE 
{\bf Given inputs}: $\mu_{p_\iter}^{[\ell]}, \sigma_{p_\iter}^{2[\ell]},z_\iter^{[\ell]},\notaI{v_\iter^{2[\ell]}},m_{\iter}^{[\ell]},\notaI{\eta}_{\iter}^{[\ell]}$
\STATE
1)  Run moment matching:  Set the mean and variance of the unnormalized Gaussian distribution
\begin{equation}\LABEQ{updategaussian}
q_E^{[\ell]}(\beforechannel_\iter)\cdot\cgauss{\beforechannel_\iter}{m_{\iter,new}^{[\ell{+1}]}}{\notaI{\eta}_{\iter,new}^{[\ell{+1}]}}
\end{equation}
equal to $\mu_{p_\iter}^{[\ell]}$ and $\sigma_{p_\iter}^{2[\ell]}$, to get the solution
\begin{align}
\notaI{\eta}_{\iter,new}^{[\ell+1]}&=\frac{\sigma_{p_\iter}^{2[\ell]}\notaI{v_\iter^{2[\ell]}}}{\notaI{v_\iter^{2[\ell]}}-\sigma_{p_\iter}^{2[\ell]}} , \LABEQ{Lambdak1new} \\
%v_{\iter,new}^{[\ell+1]}&=\left({\sigma_{p_\iter}^{-2[\ell]}}-{t_\iter^{-[\ell]}} \right)\inv , \LABEQ{Lambdak1new} \\
m_{\iter,new}^{[\ell+1]}&= \notaI{\eta}_{\iter,new}^{[\ell+1]}{\left( \frac{\mu_{p_\iter}^{[\ell]}}{\sigma_{p_\iter}^{2[\ell]}}-\frac{z_\iter^{[\ell]}}{\notaI{v_\iter^{2[\ell]}}} \right)} . 
\end{align}
\STATE
2) Run damping: Update the values as
\begin{align}
\notaI{\eta}_\iter^{[\ell+1]}&=\left(\beta\frac{1}{\notaI{\eta}_{\iter,new}^{[\ell+1]}} + (1-\beta)\frac{1}{\notaI{\eta}_\iter^{[\ell]}}\right)\inv \LABEQ{Lambdak1} ,\\
m_\iter^{[\ell+1]}&=\notaI{\eta}_\iter^{[\ell+1]}\left(\beta \frac{m_{\iter,new}^{[\ell+1]}}{\notaI{\eta}_{\iter,new}^{[\ell+1]}} + (1-\beta)\frac{m_\iter^{[\ell]}}{\notaI{\eta}_\iter^{[\ell]}}\right). \LABEQ{Lambdak2}
\end{align}
\IF{$\notaI{\eta}_{\iter}^{[\ell+1]}<0$}
\vspace{-0.3cm}
\STATE
\begin{align}
\notaI{\eta}_{\iter}^{[\ell+1]}=\notaI{\eta}_{\iter}^{[\ell]}, \,\,\,\,\,\,\,\, m_{\iter}^{[\ell+1]}=m_{\iter}^{[\ell]}. 
\end{align}
\ENDIF
\STATE
{\bf Output}: $\notaI{\eta}_{\iter}^{[\ell+1]}, m_{\iter}^{[\ell+1]}$
\end{algorithmic}
\caption{Moment Matching and Damping}\LABALG{MMD}
\end{algorithm}

%To start the algorithm, we need to initialize 

\subsection{The nuBEP algorithm}
\ALG{nuEPalgorithm} contains a detailed description of the whole EP procedure\notajj{, where $\iterep$ is the number of EP iterations while $\nturbo$ is the number of turbo iterations}. 
%A detail explanation of the algorithm is described in \ALG{nuEPalgorithm}. 
\notaI{Note that the difference with the approach in \cite{Santos16} lies in the definition of the prior distribution used during the moment matching procedure. In \cite{Santos16}, we use an uniform distribution (denoted with the indicator function), forcing the same \textit{a priori} probability for the symbols, regardless of the information fed back from the decoder, during the moment matching employed in the equalizer even after the turbo procedure starts.}
%Note that the difference with the approach in \cite{Santos16} lies in the definition of the prior distribution used. In \cite{Santos16}, we use an uniform distribution, forcing the same \notaI{\textit{a priori}} probability for the symbols regardless of the information fed back from the decoder in the turbo equalization iterations. 
In this paper, we refine the definition of the prior used \notaI{in the moment matching} of EP algorithm as in  \EQ{indfunc}, considering non-uniform priors once the turbo procedure starts. For this reason, we named this algorithm non-uniform BEP (nuBEP) turbo equalizer.

\subsection{On the election of EP parameters}\LABSSEC{parameters}

The moment matching condition explained in \EQ{MM} determines the optimal operation point found by the EP approximation. By repeating this procedure, we allow to find a stationary solution for the operation point. In order to avoid instabilities and control the accuracy and speed of convergence, some EP parameters are introduced. These parameters are the number of EP iterations ($\iterep$), a minimum allowed variance ($\epsilon$) and a damping factor ($\beta$). 
Based on recent studies, these EP parameters can be further optimized \cite{Opper05,Cespedes17thesis,Cespedes17}. Following the guidelines in those papers and after extensive experimentation, in the general case we found out that instabilities can be controlled by setting\footnote{\notaI{Parameters have been chosen to optimize turbo equalization \cite{Cespedes17}.} } $\epsilon=1e^{-8}$. %For $64$-QAM modulations, the best performance has been found by setting $\epsilon=0.9$. 
Regarding the accuracy of the algorithm, it is convenient to start with a conservative value of the \notaI{damping} parameter $\beta$ in \ALG{nuEPalgorithm}. The value $\beta=0.1$ forces our algorithm to move slowly towards the EP solution. Once the turbo procedure starts, we let the \notaI{damping} parameter grow in order to speed up the achievement of the EP solution, reducing the value of $\iterep$ from 10 in \cite{Santos16} to 3. A simple rule for determination of $\beta$ that fulfills this requirements and leads to good performance is an exponential growth with a saturation value of $0.7$, i.e., $\beta=\min(\exp^{t/1.5}/10),0.7)$, where $t\in[0,\nturbo]$ is the number of the current turbo iteration. 
With this criterion the number of EP iterations after the turbo procedure starts is reduced to $\iterep=3$, hence reducing the computational complexity by more than a third. 

%\notajj{En Algoritmo 2 me pierdo con las $p$ y las $p_D$. Las he puesto todas a $p_D$}

\begin{algorithm}[!tb]
\begin{algorithmic}
\STATE 
{\bf Initialization}: Set \notaI{$p_D(\beforechannel_\iter)=\frac{1}{\modsize}\sum_{\beforechannel\in\mathcal{A}} \delta(\beforechannel_\iter-\beforechannel)$} for $\iter=1,\hdots, \tamframe$
%Receive $p(\beforechannel_\iter)$ from the decoder for $\iter=1,\hdots, \tamframe$\\
%If $p(\beforechannel_\iter)$ is not available, set $p(\beforechannel_\iter)=\frac{1}{\modsize}$ \\
%Initialize $(m_\iter^{[1]}, v_\iter^{[1]})=(0, \energy)$. 
\FOR {$t=1,...,\nturbo$}
%\STATE
%0) If $p(\beforechannel_\iter)$ is not available, set $p(\beforechannel_\iter)=\frac{1}{\modsize}$ 
\STATE
\notaI{1) Compute the mean $m_\iter^{[1]}$ and variance $\notaI{\eta}_\iter^{[1]}$ given by \EQ{meanturbo} and \EQ{varturbo}, respectively. }
%\STATE
%2) Set $(m_\iter^{[1]}, v_\iter^{[1]})=(m_\iter,v_\iter)$. 
%\IF {t=1}
%\STATE
%Set $p(\beforechannel_\iter)=\frac{1}{\modsize}$
%\ENDIF \\
%\STATE
\FOR {$\ell=1,...,\iterep$}
\FOR {$\iter=1,...,\tamframe$}
\STATE
2) Compute the $\iter$-th extrinsic distribution as in \EQ{extrinsicnubep}, i.e., 
\begin{align}\LABEQ{cavitynuEP}
q_E^{[\ell]}(\beforechannel_\iter)= \cgauss{\beforechannel_\iter}{z_\iter^{{[\ell]}}}{\notaI{v_\iter^{{2[\ell]}}}}
\end{align}
where $z_\iter^{{[\ell]}}$ and \notaI{$v_\iter^{{2[\ell]}}$} are given by \EQ{mean_ext} and \EQ{var_ext}, respectively. 
\STATE
3)  Obtain the distribution $\widehat{p}^{[\ell]}(\beforechannel_\iter)\;\propto\;q_E^{[\ell]}(\beforechannel_\iter)p_D(\beforechannel_\iter)$ and estimate its mean $\mu_{p_\iter}^{[\ell]}$ and variance $\sigma_{p_\iter}^{2[\ell]}$. Set a minimum allowed variance as \mbox{$\sigma_{p_\iter}^{2[\ell]}=\max(\epsilon,\sigma_{{p}_\iter}^{2[\ell]})$}. 
\STATE
4)  Run the moment matching and damping procedures by executing \ALG{MMD}. 
\ENDFOR 
\ENDFOR \\
\STATE
5) With the values $m_\iter^{[\iterep+1]}, \notaI{\eta}_\iter^{[\iterep+1]}$ obtained after the EP algorithm, calculate the extrinsic distribution $q_E(\beforechannel_\iter)$. % in \EQ{extrinsicnubep} and deliver it to the channel decoder through the demapper. \\
\STATE
\notaI{6) Demap the extrinsic distribution and compute the extrinsic LLR, $L_E(c_{\iter,\iterj})$, by means of \EQ{LLRdemap}. }
\STATE
\notajj{7) Run the channel decoder to output $p_D(\beforechannel_\iter)$}
%\STATE
%5) Receive $p(\beforechannel_\iter)$ from the decoder and compute its mean $m_\iter$ and variance $v_\iter$ given by \EQ{meanturbo} and \EQ{varturbo}, respectively.
%\STATE
%6) Set $(m_\iter^{[1]}, v_\iter^{[1]})=(m_\iter,v_\iter)$. 
\ENDFOR
\STATE
\notaI{{\bf Output}: Deliver $L_E(c_{\iter,\iterj})$ to the channel decoder for $\iter=1,\hdots, \tamframe$ and $j=1,\hdots,Q$}
\end{algorithmic}
\caption{nuBEP Turbo Equalizer}\LABALG{nuEPalgorithm}
\end{algorithm}

%The value of the smoothing parameter $\beta$ has to start with a small value and increase it with the number of the turbo iterations. Specifically, we set it to $\beta=\min(\exp^{t/1.5}/10),0.7)$, where $t\in[0,\nturbo]$ is the number of the current turbo iteration. Note that before the turbo procedure $\beta=0.1$, a conservative value that forces our algorithm to move slowly towards the EP solution and needs $\iterep=10$ iterations to get an accurate enough solution. Once the turbo procedure starts, we let the smoothing parameter to grow exponentially, at a maximum value of 0.7, which allows to get the EP solution sooner and reduce the number of the EP iterations to a third part, $\iterep=3$. 

%\subsection{Differences with BEP in \cite{Santos17}}

\section{Filter-type turbo equalization}\LABSEC{Filter}
\subsection{LMMSE filter}\LABSSEC{MMSE-F}
\notaI{In this subsection we review the formulation of the LMMSE-based filter \cite{Singer01,Koetter02,Tuchler11}, modified to allow for unnormalized transmitted energy and a different computation of the extrinsic distribution. }
The LMMSE-based filter \cite{Singer01,Koetter02,Tuchler11} estimates one symbol per $\iter$-th iteration, $\beforechannel_\iter$, given a $\winsize$-size window of observations, $\vecti{\afterchannel}{\iter}=[\afterchannel_{\iter-\winsize_2}, ..., \afterchannel_{\iter+\winsize_1}]\trs$, where $\winsize=\winsize_1+\winsize_2+1$. This procedure differs from \cite{Wu07}, where each transmitted symbol is estimated given the whole vector of observations, $\vect{\afterchannel}$. 
{The LMMSE equalizer approximates the prior for each symbol, $p_D(\beforechannel_\iter)$, as a Gaussian
\begin{equation}\LABEQ{priormmse}
p_D(\beforechannel_\iter)\approx \notaI{\tilde{p}_D}(\beforechannel_\iter)=\cgauss{\beforechannel_\iter}{m_\iter}{\notaI{\eta}_\iter},
\end{equation}
where the mean, $m_\iter$, and variance, $v_\iter$, are \notaI{\textit{a priori}} statistics for each transmitted symbol, given by \EQ{meanturbo} and \EQ{varturbo}, respectively. 
%\begin{align}\LABEQ{meanturbo}
%m_\iter&=\sum_{\beforechannel\in\mathcal{A}} \beforechannel \cdot p(\beforechannel_\iter=\beforechannel), \\
%v_\iter&=\sum_{\beforechannel\in\mathcal{A}}  |\beforechannel-m_\iter|^2 \cdot p(\beforechannel_\iter=\beforechannel). \LABEQ{varturbo}
%\end{align}
%For the first iteration of the turbo equalization no a priori information is available and we set $m_\iter=0$ and $v_\iter=\energy$. 
For the first iteration of the turbo equalization no \notaI{\textit{a priori}} information is available and a suitable initialization is $m_\iter=0$, $\notaI{\eta}_\iter=\energy$, which boils down to $m_\iter=0$, $\notaI{\eta}_\iter=1$ when normalizing the energy \cite{Singer01,Koetter02,Tuchler11}. 
Given the current prior and the channel impulse response (CIR), the LMMSE filter computes a Gaussian approximation of the \emph{posterior} probability of each symbol. When a turbo scheme is used, the equalizer and decoder exchange \emph{extrinsic} information \cite{Koetter04}. Through the turbo equalization iterations,  the \notaI{\textit{a priori}} statistics in \EQ{priormmse} are updated with the information fed back from the channel decoder.}

{Rather than computing the posterior distribution as in \EQ{pugiveny2}, the LMMSE filter \cite{Koetter02} considers the a posteriori probabilities with respect to the estimated transmitted symbol, $\widehat{\beforechannel}_\iter$. For this reason, and to keep the same notation than in \cite{Koetter02}, we will denote the approximated posterior as $q(\beforechannel_\iter|\widehat{\beforechannel}_\iter)$. With this posterior distribution in mind, the extrinsic probability at the output of the LMMSE filter can be computed as
\begin{equation}\LABEQ{ext2-1}
q_E(\beforechannel_\iter|\widehat{\beforechannel}_\iter)=\frac{q(\beforechannel_\iter|\widehat{\beforechannel}_\iter)}{\notaI{\tilde{p}_D}(\beforechannel_\iter)}. 
\end{equation}
This distribution is \notajj{Gaussian} and can be derived from the extrinsic distribution of the estimated symbol computed in \cite{Koetter02}, as shown in \APEN{ap2}, yielding
\begin{equation}\LABEQ{ext2}
q_E(\beforechannel_\iter|\widehat{\beforechannel}_\iter)= \cgauss{\beforechannel_\iter}{z_\iter}{\notaI{v_\iter^2}}
\end{equation}
where
\begin{align}\LABEQ{mean_extF}
z_\iter&= \frac{\vecti{\vcoef}{\iter}\her(\vecti{\afterchannel}{\iter} - \vecti{H}{\winsize}\vecti{m}{\iter}+m_\iter\vecti{h}{\winsize})}{\vecti{\vcoef}{\iter}\her\vecti{h}{\winsize}}, \\
\notaI{v_\iter^2}&= \frac{\vecti{\vcoef}{\iter}\her\vecti{h}{\winsize}\energy(1-\vecti{h}{\winsize}\her\vecti{\vcoef}{\iter})}{(\vecti{\vcoef}{\iter}\her\vecti{h}{\winsize})^2}, \LABEQ{var_extF}
\end{align}
and, in turn, 
\begin{align}\LABEQ{vcoef}
\vecti{\vcoef}{\iter}&=\left( \vecti{\Sigma}{\iter}+(\energy-\notaI{\eta}_\iter)\vecti{h}{\winsize}\vecti{h}{\winsize}\her \right)\inv\energy\vecti{h}{\winsize}, \\
%\end{equation}
%Matrix $\vecti{H}{\winsize}$ 
%\begin{align}\LABEQ{Hk}
\vecti{H}{\winsize}&=
\begin{bmatrix}
h_\ntaps & \hdots & h_1 & & & \matr{0}\\
 & \ddots &  & \ddots  \\
& & \ddots & & \ddots\\
\matr{0}& & & h_{\ntaps} & \hdots & h_1\\
\end{bmatrix}\LABEQ{Hk}
\end{align}
is the $\winsize\times (\winsize+\ntaps-1)$ channel matrix, 
$\vecti{h}{\winsize}$ is the ($\winsize_2+\ntaps$)-th column of $\vecti{H}{\winsize}$ %, $m_\iter$ and $v_\iter$ are the a priori mean and variance of the transmitted symbol $\beforechannel_\iter$, respectively, 
and
\begin{align}
\vecti{m}{\iter}&=[m_{\iter-\ntaps-\winsize_2+1}, ..., m_{\iter+\winsize_1}]\trs, \\
\vecti{V}{\iter}&=\mbox{diag}(\notaI{\eta}_{\iter-\ntaps-\winsize_2+1}, ..., \notaI{\eta}_{\iter+\winsize_1}), \\
\vecti{\Sigma}{\iter}&=\sigma_\noise^2\matr{I}+\vecti{H}{\winsize} \vecti{V}{\iter}\vecti{H}{\winsize}\her. 
\end{align}
The computational complexity is dominated by \EQ{vcoef}, which has to be recomputed every $\iter$-th iteration. Hence, the complexity is $\order(\tamframe\winsize^2)$. This complexity can be further reduced by relying on some approximations proposed in \cite{Koetter02,Tuchler11}. }

\subsection{EP filter (EP-F)}\LABSSEC{EP-F}

%EP \cite{Minka01thesis,Minka01,Seeger05,Bishop06} is a technique in Bayesian machine learning that approximates a (non-exponential) distribution with an exponential distribution whose moments match the true ones. In this paper, we focus on computing a Gaussian approximation for the posterior in \EQ{pugiveny}, which is clearly non Gaussian due to the product of indicator functions. 
\notaIc{A novel EP filter-type is developed in this subsection to improve the accuracy and performance of the LMMSE-based filter explained above. }
As explained in \SSEC{MMSE-F}, if the LMMSE filter is run, the prior of each symbol is approximated by a Gaussian with the statistics given by the decoder, i.e., with mean and variance given by \EQ{meanturbo} and \EQ{varturbo}, respectively. %This yields the posterior in \EQ{posMMSE}. 
By using the EP algorithm \notajj{we approximate the posterior distribution with a Gaussian family. Since the posterior distribution includes the true discrete priors, we take into account the discrete nature of symbols.} %keeping the true prior of the different symbols in turns and finding the Gaussian posterior whose moments match those of this new posterior distribution. 

%\notaI{Following a similar procedure than in \SEC{nuBEP}, at every EP iteration, we replace the indicator functions in \EQ{pugiveny} by gaussians in order to find a gaussian approximation to the true posterior. As explained in the LMMSE filter, rather than compute the posterior as \EQ{pugiveny}, we will compute the probability with respect to the estimated transmitted symbol, yielding an extrinsic distribution as }

At every iteration of the EP algorithm, $\ell$, we \notajj{approximate} the product of priors of individual symbols in \EQ{pugiveny2} as a product of $\tamframe$ Gaussians,   $\notaI{\tilde{p}_D^{[\ell]}}(\beforechannel_\iter)=\cgauss{\beforechannel_\iter}{m_\iter^{[\ell]}}{\notaI{\eta}_\iter^{[\ell]}}$, whose parameters (means and variances) are adjusted  to find a better approximation, $q^{[\ell]}(\vect{\beforechannel}) \; \propto \; p(\vect{\afterchannel}|\vect{\beforechannel}) \prod_{\iter=1}^{\tamframe}\notaI{\tilde{p}_D^{[\ell]}}(\beforechannel_\iter)$, to the true posterior. %Let us denote the current Gaussian prior by, where the index $\ell$ indicates the iteration of the EP algorithm. 
Similarly to \EQ{ext2-1}-\EQ{ext2}, for each $\iter$-th symbol, we first compute the current extrinsic distribution, % as % by removing all prior information about $\beforechannel_\iter$ from the current posterior, i.e., 
\begin{equation}\LABEQ{extEP}
q_E^{[\ell]}(\beforechannel_\iter|\widehat{\beforechannel}_\iter)=\frac{q^{[\ell]}(\beforechannel_\iter|\widehat{\beforechannel}_\iter)}{\notaI{\tilde{p}_D^{[\ell]}}(\beforechannel_\iter)}.
\end{equation}
%\notajj{NOTACION:Esto no tiene mucho sentido. Se usa $q$ para denotar la marginal de la aproximacion y ahora para denotar el producto de la cavity por el prior verdadero :-0. Ademas en el texto queda raro repetir (37) $\rightarrow$ approximation to (37) with same moments, as in (15).} 
\notajj{Now, a more accurate posterior distribution can be obtained by finding \notaI{a new} Gaussian approximation, $\tilde{p}_D^{[\ell+1]}(\beforechannel_\iter)$, \notaI{to match the moments of $q_E^{[\ell]}(\beforechannel_\iter|\widehat{\beforechannel}_\iter) \tilde{p}_D^{[\ell+1]}(\beforechannel_\iter)$ and $q_E^{[\ell]}(\beforechannel_\iter|\widehat{\beforechannel}_\iter) p_D(\beforechannel_\iter)$}, as in \EQ{MM}. }
%
%
%multiplying the previous extrinsic distribution by the true prior,
%\begin{equation}\LABEQ{posep}
%\notaI{\widehat{p}^{[\ell]}(\beforechannel_\iter)\;\propto\;} q_E^{[\ell]}(\beforechannel_\iter|\widehat{\beforechannel}_\iter) p_D(\beforechannel_\iter),
%\end{equation}
%which is clearly a non-Gaussian distribution. Finally, we find the Gaussian approximation to \EQ{posep} whose moments match those of \EQ{posep}, as in \EQ{MM}. This yields 
%\begin{align}\LABEQ{epSantos}
%\notaI{\tilde{p}_D^{[\ell+1]}}(\beforechannel_\iter)
%=\cgauss{\beforechannel_\iter}{m_\iter^{[\ell+1]}}{\notaI{\eta}_\iter^{[\ell+1]}}.
%\end{align}
With these new values for the mean, $m_\iter^{[\ell+1]}$,  and variance, $\notaI{\eta}_\iter^{[\ell+1]}$, we can recompute a new extrinsic distribution $q_E^{[\ell+1]}(\beforechannel_\iter|\widehat{\beforechannel}_\iter)$, which is more accurate than the one in \EQ{ext2}. {The final extrinsic distribution delivered to the decoder is the one obtained after the last iteration of the EP algorithm, following \EQ{extEP}. }

\begin{algorithm}[!tb]
\begin{algorithmic}
\STATE
%{\bf Given input}: Receive $p_D(\beforechannel_\iter)$ from the decoder for $\iter=1,\hdots, \tamframe$\\
{\bf Initialization}: Set \notaI{$p_D(\beforechannel_\iter)=\frac{1}{\modsize}\sum_{\beforechannel\in\mathcal{A}} \delta(\beforechannel_\iter-\beforechannel)$} for $\iter=1,\hdots, \tamframe$
%\STATE
%Initialize $(m_\iter^{[1]}, v_\iter^{[1]})=(0, \energy)$. 
\FOR {$t=1,...,\nturbo$}
\STATE
%0) If $p(\beforechannel_\iter)$ is not available, set $p_D(\beforechannel_\iter)=\frac{1}{\modsize}$ 
%\STATE
1) Compute the mean \notaI{$m_\iter^{[1]}$} and variance \notaI{$\notaI{\eta}_\iter^{[1]}$} given by \EQ{meanturbo} and \EQ{varturbo}, respectively. 
%\STATE
%2) Set $(m_\iter^{[1]}, v_\iter^{[1]})=(m_\iter,v_\iter)$. 
%\STATE
\FOR {$\ell=1,...,\iterep$}
%\STATE
%Calculate the distribution $\aproxfunc^{[\ell]}(\vect{\beforechannel})$ in \EQ{aproxepblock} with  $\gamma_\iter\leftarrow\gamma_\iter^{[\ell]}$ and $\Lambda_\iter\leftarrow\Lambda_\iter^{[\ell]}$.
\FOR {$\iter=1,...,\tamframe$}
\STATE
2) Compute the $\iter$-th extrinsic distribution as in \EQ{ext2}, i.e., 
\begin{align}\LABEQ{cavityEP}
q_E^{[\ell]}(\beforechannel_\iter|\widehat{\beforechannel}_\iter)= \cgauss{\beforechannel_\iter}{z_\iter^{{[\ell]}}}{\notaI{v_\iter^{{2[\ell]}}}}
\end{align}
where $z_\iter^{{[\ell]}}$ and \notaI{$v_\iter^{{2[\ell]}}$} are given by \EQ{mean_extF} and \EQ{var_extF}, respectively. 
%\begin{align}
%z_\iter^{{[\ell]}}&= \frac{\vecti{\vcoef}{\iter}\trs(\vecti{\afterchannel}{\iter} - \vecti{H}{\iter}\vecti{m}{\iter}^{[\ell]}+m_\iter^{[\ell]}\vecti{h}{\iter})}{\vecti{\vcoef}{\iter}\trs\vecti{h}{\iter}}, \\
%t_\iter^{{[\ell]}}&= \frac{\vecti{\vcoef}{\iter}\trs\vecti{h}{\iter}\energy(1-\vecti{h}{\iter}\trs\vecti{\vcoef}{\iter})}{(\vecti{\vcoef}{\iter}\trs\vecti{h}{\iter})^2}, \\
%\vecti{\vcoef}{\iter}&=\left( \vecti{\Sigma}{\iter}+(\energy-v_\iter^{[\ell]})\vecti{h}{\iter}\vecti{h}{\iter}\trs \right)\inv\energy\vecti{h}{\iter}.
%\end{align}
\STATE
3)  Obtain the distribution $\widehat{p}^{[\ell]}(\beforechannel_\iter)\;\propto\;q_E^{[\ell]}(\beforechannel_\iter|\widehat{\beforechannel}_\iter)p_D(\beforechannel_\iter)$ and estimate its mean $\mu_{p_\iter}^{[\ell]}$ and variance $\sigma_{p_\iter}^{2[\ell]}$. Set a minimum allowed variance as \mbox{$\sigma_{p_\iter}^{2[\ell]}=\max(\epsilon,\sigma_{{p}_\iter}^{2[\ell]})$}. 
\STATE
4)  Run the moment matching and damping procedures by executing \ALG{MMD}. 
%Moment matching: Set the mean and variance of the unnormalized Gaussian distribution
%\begin{equation}\LABEQ{updategaussian}
%q_E^{[\ell]}(\beforechannel_\iter|\widehat{\beforechannel}_\iter)\cdot\gauss{\beforechannel_\iter}{m_{\iter,new}^{[\ell\notaI{+1}]}}{v_{\iter,new}^{[\ell\notaI{+1}]}}
%\end{equation}
%equal to $\mu_{p_\iter}^{[\ell]}$ and $\sigma_{p_\iter}^{2[\ell]}$, to get the solution
%\begin{align}
%v_{\iter,new}^{[\ell+1]}&=\frac{\sigma_{p_\iter}^{2[\ell]}t_\iter^{[\ell]}}{t_\iter^{[\ell]}-\sigma_{p_\iter}^{2[\ell]}} , \LABEQ{Lambdak1new} \\
%%v_{\iter,new}^{[\ell+1]}&=\left({\sigma_{p_\iter}^{-2[\ell]}}-{t_\iter^{-[\ell]}} \right)\inv , \LABEQ{Lambdak1new} \\
%m_{\iter,new}^{[\ell+1]}&= v_{\iter,new}^{[\ell+1]}{\left( \frac{\mu_{p_\iter}^{[\ell]}}{\sigma_{p_\iter}^{2[\ell]}}-\frac{z_\iter^{[\ell]}}{t_\iter^{[\ell]}} \right)} . 
%\end{align}
%\IF{$v_{\iter,new}^{[\ell+1]}<0$}
%\vspace{-0.3cm}
%\STATE
%\begin{align}
%v_{\iter,new}^{[\ell+1]}=v_{\iter,new}^{[\ell]}, \,\,\,\,\,\,\,\, m_{\iter,new}^{[\ell+1]}=m_{\iter,new}^{[\ell]}. 
%\end{align}
%\ENDIF
%\STATE
%4) Update the values as
%\begin{align}
%v_\iter^{[\ell+1]}&=\left(\beta\frac{1}{v_{\iter,new}^{[\ell+1]}} + (1-\beta)\frac{1}{v_\iter^{[\ell]}}\right)\inv \LABEQ{Lambdak1} ,\\
%m_\iter^{[\ell+1]}&=v_\iter^{[\ell+1]}\left(\beta \frac{m_{\iter,new}^{[\ell+1]}}{v_{\iter,new}^{[\ell+1]}} + (1-\beta)\frac{m_\iter^{[\ell]}}{v_\iter^{[\ell]}}\right). \LABEQ{Lambdak2}
%\end{align}
\ENDFOR 
\ENDFOR \\
\STATE
5) With the values $m_\iter^{[\iterep+1]}, \notaI{\eta}_\iter^{[\iterep+1]}$ obtained after the EP algorithm, calculate the extrinsic distribution $q_E(\beforechannel_\iter|\widehat{\beforechannel}_\iter)$ in \EQ{ext2}. \\% and deliver it to the channel decoder through the demapper. \\
%$\boldsymbol{\upmu}$ and 
%$\matr{\Sigma}$ given by \EQ{meanEP} and \EQ{covEP}, respectively. \\
%
\STATE
\notaI{6) Demap the extrinsic distribution and compute the extrinsic LLR, $L_E(c_{\iter,\iterj})$, by means of \EQ{LLRdemap}. }
\STATE
\notajj{7) Run the channel decoder to output $p_D(\beforechannel_\iter)$}
%\STATE
%5) Receive $p(\beforechannel_\iter)$ from the decoder and compute its mean $m_\iter$ and variance $v_\iter$ given by \EQ{meanturbo} and \EQ{varturbo}, respectively.
%\STATE
%6) Set $(m_\iter^{[1]}, v_\iter^{[1]})=(m_\iter,v_\iter)$. 
\ENDFOR
\STATE
\notaI{{\bf Output}: Deliver $L_E(c_{\iter,\iterj})$ to the channel decoder for $\iter=1,\hdots, \tamframe$ and $j=1,\hdots,Q$}
\end{algorithmic}
\caption{EP-F}\LABALG{EPalgorithm}
\end{algorithm}

We denote this new algorithm as EP-filter (EP-F). \ALG{EPalgorithm} is a detailed description of its implementation. Note that the main difference between \ALG{nuEPalgorithm} and \ALG{EPalgorithm} lies in the computation of the extrinsic distribution, i.e., equations \EQ{cavitynuEP} and \EQ{cavityEP}. 
%At this point it is important to remark that in the approximation proposed we retain all the knowledge on the systems by including the first factor in \EQ{aproxepblock} while approximating with EP the unknowns.
%The parameter update in \EQ{Lambdak1new} may return a negative value $v_{\iter,new}^{[\ell+1]}$ for some $\iter$'s which means that there is no pair $(m_{\iter,new}^{[\ell+1]},v_{\iter,new}^{[\ell+1]})$ that sets the variance of the Gaussian in \EQ{updategaussian} to $\sigma_{p_\iter}^{2[\ell]}$. For those $\iter$'s, we keep the values from the previous iteration.
%We introduce a smoothing parameter $\beta\in[0,1]$ and a small constant $\epsilon$. To avoid numerical instabilities, constant $\epsilon$ is the minimum allowed variance at each iteration, i.e., \mbox{$\sigma_{p_\iter}^{2[\ell]}=\max(\epsilon,\sigma_{{p}_\iter}^{2[\ell]})$}. 
The computational complexity is also dominated by \EQ{vcoef}, which has to be computed for each symbol and each $\ell$-th iteration. Hence, the complexity is $\iterep$ times the LMMSE complexity, i.e. $\order(\iterep\tamframe\winsize^2)$, where $\iterep$ is the number of iterations of the EP-F. At this point, it is interesting to remark that the approximations proposed in \cite{Koetter02,Tuchler11} to further reduce the complexity cannot be applied when the EP is used. {The reason is that these approximations remove (at some points) the prior variance computed by the decoder, setting it to one.} %However, the EP algorithm cannot obtain a more accurate for the variance wether its previous value is not available. 

%In \TAB{complexity} we include a detailed comparison of the complexity of the proposed nuBEP and EP-F, the algorithm in \cite{Santos15} (BEP), the block and filter implementation of the LMMSE and BCJR approaches, where $\Me$ is the number of survivor states and $\winsize$ depends linearly with $\ntaps$. 

%\section{Computational complexity}

\section{\notaIc{Relation to previous approaches}}
%\notaIc{This section is devoted to fully clarify which are the main contributions of the manuscript and describe the differences with existing equalizers based on the EP algorithm. The novelties can be divided into three lines: }
\subsection{\notaIc{Update of the priors} }

\notaIc{We improve the prior information used in the equalizer once the turbo procedure starts, forcing the true discrete prior to be non-uniform in contrast to the uniform priors used by previous EP-based approaches. }

\notaIc{In previous proposals \cite{Santos15,Santos16,Santos17}, the probabilities from the channel decoder, $p_D(\beforechannel_\iter)$, were used to initialize, at the beginning of every iteration of the turbo-equalization,  the product of Gaussians that in the EP approximation replaces the product of priors, $\tilde{p}_D^{[1]}(\beforechannel_\iter)$. 
%\begin{align}
%\tilde{p}_D^{[1]}(\beforechannel_\iter)=\mbox{Proj}_G(p_D(\beforechannel_\iter))
%\end{align}
But when the moment matching was performed in the EP algorithm, i.e.,
\begin{equation}\LABEQ{MM2}
q_E^{[\ell]}(\beforechannel_\iter)\mathbb{I}_{\beforechannel_\iter\in\mathcal{A}} \stackrel{\mbox{\begin{tabular}{c}moment\\matching\end{tabular}}}{\longleftrightarrow} q_E^{[\ell]}(\beforechannel_\iter) \tilde{p}_D^{[\ell+1]}(\beforechannel_\iter),
\end{equation}
the true priors used were uniformly distributed following
\begin{equation}\LABEQ{prior1}
%p_k(\beforechannel_\iter)=
\mathbb{I}_{\beforechannel_\iter\in\mathcal{A}}=\frac{1}{\modsize}\sum_{\beforechannel\in\mathcal{A}} \delta(\beforechannel_\iter-\beforechannel).
\end{equation}
In the current proposal, we keep the initialization of the Gaussians in every step of the turbo-equalization, $\tilde{p}_D^{[1]}(\beforechannel_\iter)$, but also \textit{propose to replace the uniform priors in \EQ{prior1} by {\it non-uniform} ones} in the moment matching step, as explained in \EQ{MM}, i.e., 
\begin{equation}\LABEQ{MM2}
q_E^{[\ell]}(\beforechannel_\iter)p_D(\beforechannel_\iter) \stackrel{\mbox{\begin{tabular}{c}moment\\matching\end{tabular}}}{\longleftrightarrow} q_E^{[\ell]}(\beforechannel_\iter) \tilde{p}_D^{[\ell+1]}(\beforechannel_\iter),
\end{equation}
where
 \begin{equation}\LABEQ{prior2}
%p_k(\beforechannel_\iter)=
p_D(\beforechannel_\iter)=\sum_{\beforechannel\in\mathcal{A}} \delta(\beforechannel_\iter-\beforechannel) \prod_{\iterj=1}^{Q} p_D(c_{\iter,\iterj}=\varphi_\iterj(\beforechannel)),
\end{equation}
Note that the different definition of priors -\EQ{prior1} in previous proposals, \EQ{prior2} in this manuscript- is the difference between the currently proposed nuBEP algorithm and the BEP in \cite{Santos16}, with remarkable improvements.
%Note that the definition of priors in \EQ{prior1} used in previous proposals is different to the ones in \EQ{prior2} used in this manuscript. This is the remarkable difference between the currently proposed nuBEP algorithm and the BEP in \cite{Santos16}, 
%This is the major reason why the nuBEP has remarkable improvement with respect to the previous algorithms, as will be shown in \SEC{sim}. 
}

\subsection{\notaIc{Parameter Optimization}}
 \notaIc{The computational complexity of the EP algorithm is reduced to roughly a third part of that in \cite{Santos16}, by optimizing the choice of EP parameters.}
 \notaIc{In particular, we propose some new values for $\epsilon$ and $\beta$, that control numerical instabilities in the EP updates, and $\iterep$, the number of iterations of the EP equalizer. The parameters proposed in this paper reduce the number of iterations in turbo equalization to $\iterep=3$, rather than the $\iterep=10$ iterations that were used in \cite{Santos16}. }
 
%\notaIc{The parameters involved in the EP procedure have an impact on the convergence of the algorithm. The parameters $\epsilon$ and $\beta$ control numerical instabilities in the EP updates and $\iterep$ determines the computational complexity of our algorithm. When turbo-equalization is used, a suitable tradeoff between the number of iterations of the EP and the turbo can speed up the overall convergence. 
%%For this reason, we modified the parameters proposed in previous approaches and optimized them as explained in \SSEC{parameters}, 
%The parameters proposed in this paper reduce the number of iterations in turbo equalization to $\iterep=3$, rather than the $\iterep=10$ iterations that were used in \cite{Santos16}. }

\subsection{\notaIc{Filter-type solution}}
\notaIc{ The new filter-type EP solution proposed is constrained to have linear complexity in the frame length and quadratic in the filter length, i.e., it is endowed with the same complexity order than the LMMSE filter. This complexity is not quadratic with the block length as the one of the BEP \cite{Santos16} nor cubic with the window length as complexity of the SEP \cite{Santos17}.}

%\notaIc{The core of the proposal is Algorithm 3. In previous proposals, we have analyzed the EP as a block wise solution (BEP method in \cite{Santos16}) and as a Kalman-type approach (SEP equalizer in \cite{Santos17}). However, the BEP did have a quadratic complexity with the block length while the SEP has a complexity of cubic order with the window length. In this paper we rewrite the filter-type LMMSE turbo equalizer in \cite{Singer01,Koetter02,Tuchler11} to include the benefits of the EP equalizer in previous proposals and the new improvements discussed above in this section. The result is a solution with quadratic complexity in the window-length and linear in the frame length with quite remarkable performance.}

\subsection{\notaIc{Equalization solved with EP}}

\notaIc{Regarding the EP-based equalizers proposed by other authors, the approach in \cite{Sun15,Hu06} should be mentioned. These proposals deal just with how to pass information between the channel decoder and the LMMSE equalizer. Our proposal first focuses on the EP based equalization, performed independently of the turbo iterations. %As a result, the algorithm exhibit better performance for higher order constellations. 
Therefore the approaches are quite different. Issues such as how to use the priors in the moment matching within the EP equalizer or the damping do not arise in these proposals where the improvement is related only to the handling of probabilities between blocks.}

% The authors introduce EP  to deal with discrete messages coming from the decoder (that feeds back the equalizer) and translate them into Gaussian distributed priors. Then, the LMMSE algorithm is run with these priors. In a nutshell, the EP algorithm is applied \textit{after} the channel decoder to approximate its output discrete messages as Gaussians. Our proposal focuses on the design of an equalizer, then on how to fit it into a turbo-equalization approach. These proposals in \cite{Sun15,Hu06} deal just with how to pass information between the channel decoder and the LMMSE equalizer. Therefore the approaches are quite different. If no turbo iteration is performed they do not get any improvement while our proposal does. }

\section{\notaIc{Simulation} results}\LABSEC{sim}

\begin{figure*}[htb]
\centering
\includegraphics[width=7.0625in]{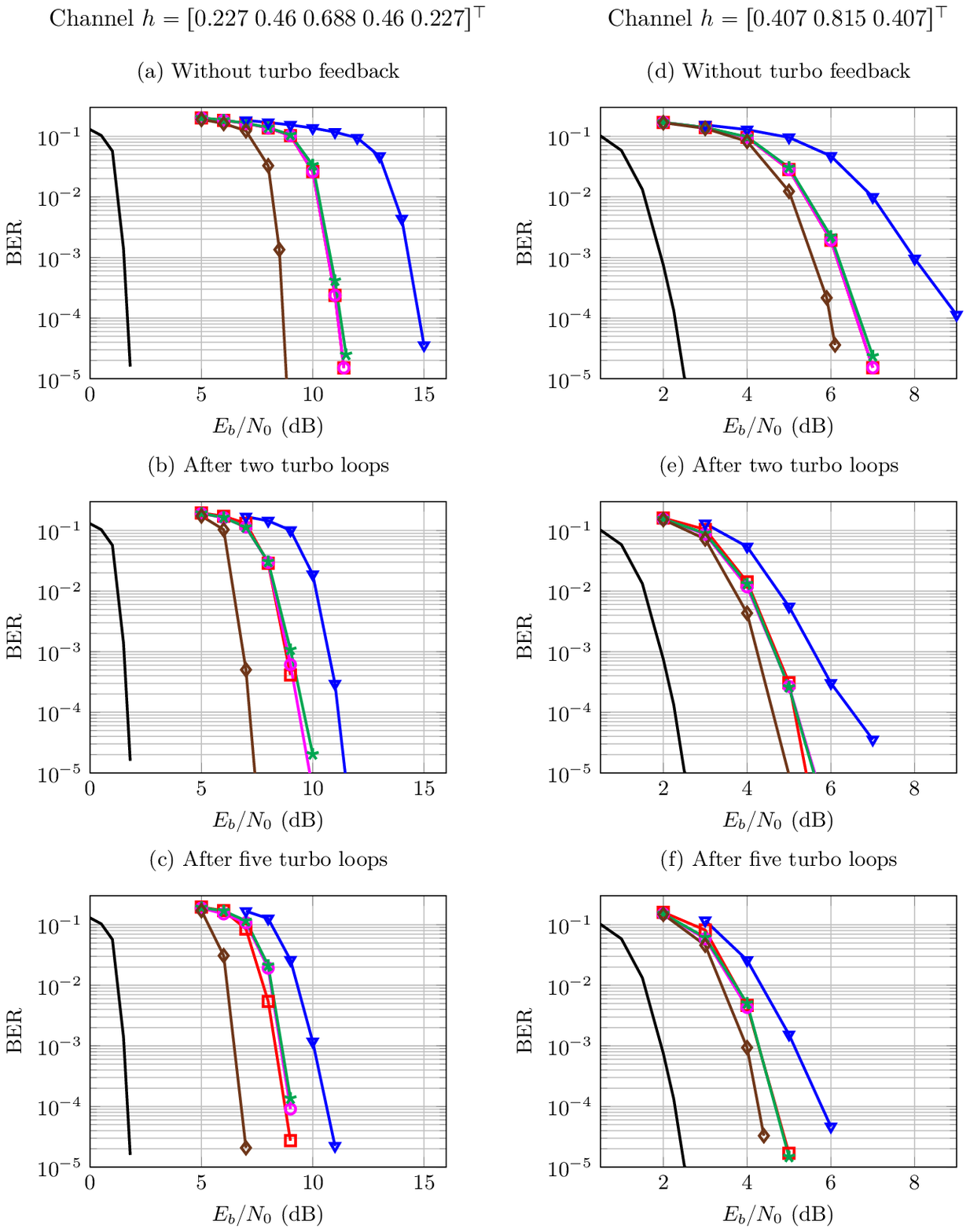}
%\scalebox{0.9}{\input{pics/BERBPSKeps1e-8b9b11.tex}}
\caption{\small BER along $\EbNo$ for BEP \cite{Santos16} (\textcolor{red}{$\square$}), nuBEP (\textcolor{mycolor3}{$\circ$}), EP-F (\textcolor{mycolor}{$\ast$}), block-LMMSE (\textcolor{blue}{$\smalltriangledown$}) and BCJR (\textcolor{auburn}{$\diamond$}) turbo equalizers, BPSK, codewords of $\t=4096$ bits (a)-(c) and $\t=1024$ bits (d)-(f) and two different channel responses. Black lines represent the AWGN bound. }\LABFIG{BPSK}
\end{figure*} 

In this section, we compare the performance of both the block LMMSE and EP-F equalizers for different scenarios. We also include the performance of the BEP \cite{Santos16} and the AWGN bound as references. 
Note that the MMSE filter \cite{Koetter02} has not been included in the simulations since the block LMMSE  exhibits equal or better performance than any filtering approaches based on the LMMSE algorithm. We did not include the SEP algorithm since it exhibits the same performance as the block implementation, as shown in \cite{Santos17}. We also include the nuBEP approach to illustrate the quite improved behavior when using non-uniform priors at each EP iteration, even reducing from 10 to 3 the number of iterations of the EP approach. The EP parameters have been selected as explained in \SSEC{parameters}, both for the nuBEP and EP-F methods. 
For a full performance comparison with BCJR approximations, such as \mbox{M-BCJR} \cite{franz98}, \mbox{M*-BCJR} \cite{Sikora05}, RS-BCJR \cite{Colavolpe01}, NZ and NZ-OS \cite{Fertonani07}, please see \cite{Santos16}. \notaI{In \TAB{complexity} we include a detailed comparison of the complexity of all the simulated algorithms. Above we include the computational complexity of previous algorithms in \cite{Santos16} (BEP) and \cite{Santos17} (SEP), the block and filter implementation of the LMMSE and BCJR approaches. Below we provide the complexity for the new approaches in this paper, i.e., the proposed nuBEP and EP-F. Parameter $\winsize$ is typically around two times the length of the channel, $\ntaps$.}
%\notajj{He cambiado orden en la tabla para poner los antiguos primero, ordenados por carga creciente, y los de este paper al final...No se si meter el SEP que sería algo así como $10N(2L+1)^3\approx80NL^3$}
\begin{table}[tb!]
\begin{center}
\captionof{table}{\small Complexity comparison between algorithms.}\LABTAB{complexity}
\includegraphics[width=3.5in]{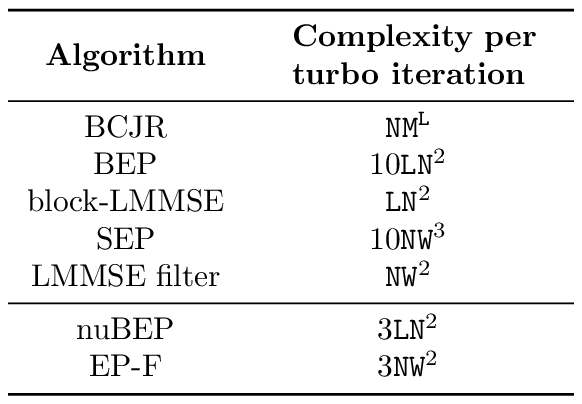}
%\captionof{\small Complexity comparison between algorithms.} \LABFIG{numcomplexity}
\end{center}
\end{table}
%
%both LMMSE \cite{Koetter02} and EP-F equalizers for different scenarios. 
%We also include the block LMMSE and BEP \cite{Santos16} performances as reference. 
%
Here, we simulate the scenarios in \cite{Singer01,Tuchler11}, using the same channel responses and modulations. Other modulations are also considered. 
% the channels whose CSI is given by $\vect{h}=[0.227\;0.46\;0.688\;0.46\;0.227]\trs$ and $h=[0.407 \; 0.815 \; 0.407]\trs$ since they are the ones shown in \cite{Singer01,Tuchler11}, respectively. 
The absolute value of LLRs given to the decoder is limited to 5 in order to avoid very confident probabilities. % which negatively affect its estimations. 
We use a (3,6)-regular LDPC of rate 1/2, and {belief propagation as decoder with a maximum of 100 iterations}. % using the belief propagation as decoder \cite{Urbanke08-2, Salamanca15}.  The codes were generated using the progressive edge-growth algorithm \cite{Hu05}.  
%The EP parameters are set to $\epsilon=1e^{-8}$, $\beta=0.1$ and $\iterep=10$. %In order to avoid instabilities for large constellations, we fix $\epsilon=0.5$ (just before the turbo procedure) when $\modsize>2$. 
%These values has been chosen after extensive experimental results. 
%$\epsilon=0.5$ 
%2^{-\max(\ell-5,1)}$ 
%(before the turbo procedure), $\epsilon=1e^{-8}$ (in the turbo procedure), $\beta=0.1$ and $\iterep=10$. 
The window length in the filtered approach is set to $\winsize=\winsize_1+\winsize_2+1$, where $\winsize_1=2\ntaps$ and $\winsize_2=\ntaps+1$ as suggested in \cite{Tuchler11}. 

In the following, we first include a section to analyze the performance of our approach in a low complexity scenario with BPSK modulation, similarly to \cite{Tuchler11}. The optimal BCJR algorithm can be run in this scenario with a low enough computational complexity and is used as bound. Next, we include a section to analyze the behavior of the algorithms in a large complexity scenario, where we use high-order modulations such as 8-PSK, 16-QAM and 64-QAM. %The BCJR algorithm cannot be run due to its large complexity and the only feasible equalizers are the ones based on the LMMSE approach or our solutions.  

\subsection{BPSK scenario}

In \FIG{BPSK} we {include} the BER, averaged {over $10^4$ random frames}, for the LMMSE, 
%LMMSE filter \cite{Koetter02}, 
BEP \cite{Santos16}, nuBEP, EP-F and BCJR equalizers with a BPSK modulation and two different channel responses and lengths of encoded words: $\vect{h}=[0.227\;0.46\;0.688\;0.46\;0.227]\trs$ and $\t=4096$ bits in \FIG{BPSK} (a)-(c) and $h=[0.407 \; 0.815 \; 0.407]\trs$ and $\t=1024$ bits in \FIG{BPSK} (d)-(f). The channel responses 
%and codeword lengths 
were selected following the simulations in \cite{Singer01,Tuchler11}. %\notaI{En realidad esta frase no es 100\% cierta porque (a)-(c) se simulan en el paper de Tuchler con palabras de 32768, pero para nosotros simular eso es inviable. }
%  and encoded words of length $\t=1024$ bits. Dashed lines represent the performance with simple {equalization and channel decoding} (no feedback from the decoder), while solid lines represent the fifth iteration of the turbo scheme. 
 The performances of block-algorithms, BEP and nuBEP, are very similar to the equivalent forward filtering approach. %, but with quadratic complexity in $\tamframe$. 
When the nuBEP algorithm is applied, 2 and 1.5 dBs {gains} are obtained compared to LMMSE approach in the turbo scenario, for the two simulated scenarios, respectively.  The EP-F exhibits a  performance similar to that of the nuBEP. %While BP-EP presents instabilities at large $\EbNo$, BEP algorithm is robust due to its damping mechanism explained in \SSEC{damp}. 
%The same scenario is simulated in \FIG{BPSK}(b) using the channel response $h=[0.407 \; 0.815 \; 0.407]\trs$, also used in \cite{Tuchler11}. %, with similar good results. 
%%

%\begin{figure}[htb]
%\centering
%\scalebox{0.95}{\input{pics/EXITeps1e-8New.tex}}
%\caption{\small EXIT charts for the decoder, the BEP \cite{Santos16}, nuBEP, EP-F and LMMSE equalizers with $7$ (dashed) and $9$ (solid) dB of $\EbNo$, BPSK modulation and codewords of $\t=1024$. } \LABFIG{EXIT}
%\end{figure} 

\begin{figure}[htb]
\centering
\includegraphics[width=3.5in]{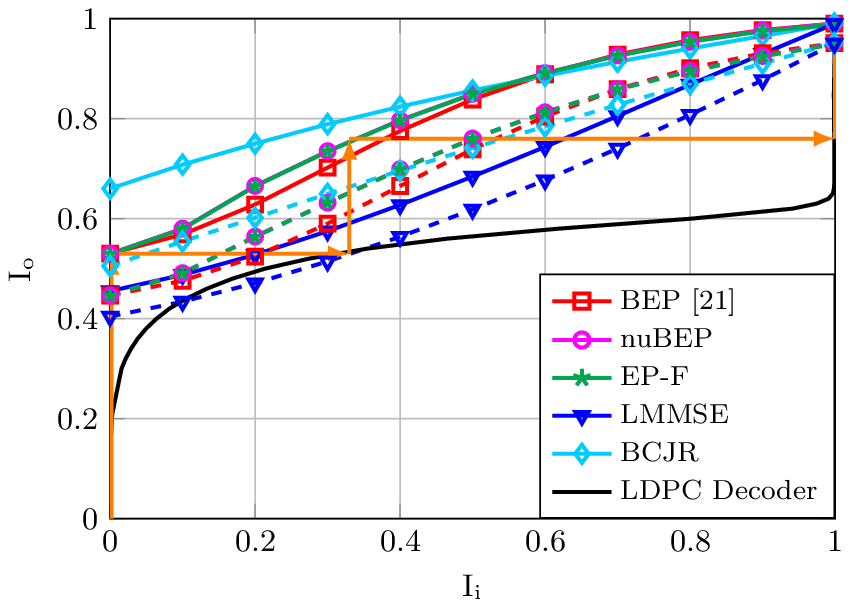}
%\scalebox{0.95}{\input{pics/EXITeps1e-8Newb11BCJR.tex}}
\caption{\small EXIT charts for the decoder, the BEP \cite{Santos16}, nuBEP, EP-F, LMMSE and \notaI{BCJR} equalizers with $7$ (dashed) and $9$ (solid) dB of $\EbNo$, BPSK modulation and codewords of \notaI{$\t=4096$}. } \LABFIG{EXIT}
\end{figure} 

In \FIG{EXIT} we include the EXIT charts of the BEP \cite{Santos16}, nuBEP, EP-F, LMMSE and \notaI{BCJR}
%and LMMSE-F (\textcolor{mycolor2}{$\circ$}) 
for the channel response $\vect{h}=[0.227\;0.46\;0.688\;0.46\;0.227]\trs$ as in \cite{Singer01, Koetter02}, BPSK modulation with $\EbNo=9$ (solid) and $7$ dB (dashed). %We used the previous channel response because is the one used in \cite{Singer01, Koetter02}. 
The EXIT chart of the LDPC encoder of \notaI{$2048/4096$} and $\rate=1/2$ used is also depicted (solid). The horizontal and vertical axis depict the mutual information at the input, $I_i$, and the output, $I_o$, respectively. 
%The horizontal axis is the mutual information of the LLR at the input, $I_i$. The vertical axis is the same measure at the output, $I_o$. 
We use arrows to show the {evolution of the mutual information along} the turbo iterations {for} $E_b/N_o=9$ dB. 
Vertical (horizontal) arrows indicate {the improvement in the mutual information} each time the equalizer (channel decoder) is executed. 
%Note that $I_o$ at the output of the equalizer is also the value at the input of the channel decoder. 
When no \notaI{\textit{a priori}} information is given to the decoder, i.e., $I_i=0$, both BEP and EP-F {provide} a higher value for the mutual information at the output, $I_o$, than the LMMSE approach, i.e., they start from a more accurate {estimation} even before the turbo equalization. {This greatly improves  the performance as it enlarges the gap between the equalizer and the channel decoder EXIT curves.} It can be seen that the LMMSE approach will fail when $E_b/N_o=7$ dB, because {both curves intersect. %It is also predicted a similar performance for the filters and block approaches for $7$ and $9$ dB of $E_b/N_o$.} % andshown that the performance obtained with both LMMSE and EP filters is exactly the same than the one got with block approaches. 
%Since the EXIT chart for the BEP and $\EbNo=4$ dB is close to the one of the LMMSE for $\EbNo=6$ dB, a 2 dB gain is expected in this scenario. %{Quitamos el MMSE?}

\notaIc{Note that the wide EXIT tunnel from the equalizer to the LDPC decoder is suggesting that the code is not optimum in terms of capacity  \cite{Xie13}.  An optimal code in this sense would exhibit an EXIT chart near below the one of the equalizer and above the curve of the $1/2$ rate LDPC code used. The design of this code for the channel equalization response is out of the scope of this paper and remains as a future line of research. }

\subsection{Large complexity scenario}

\begin{figure*}[htb]
\centering
\includegraphics[width=7.0625in]{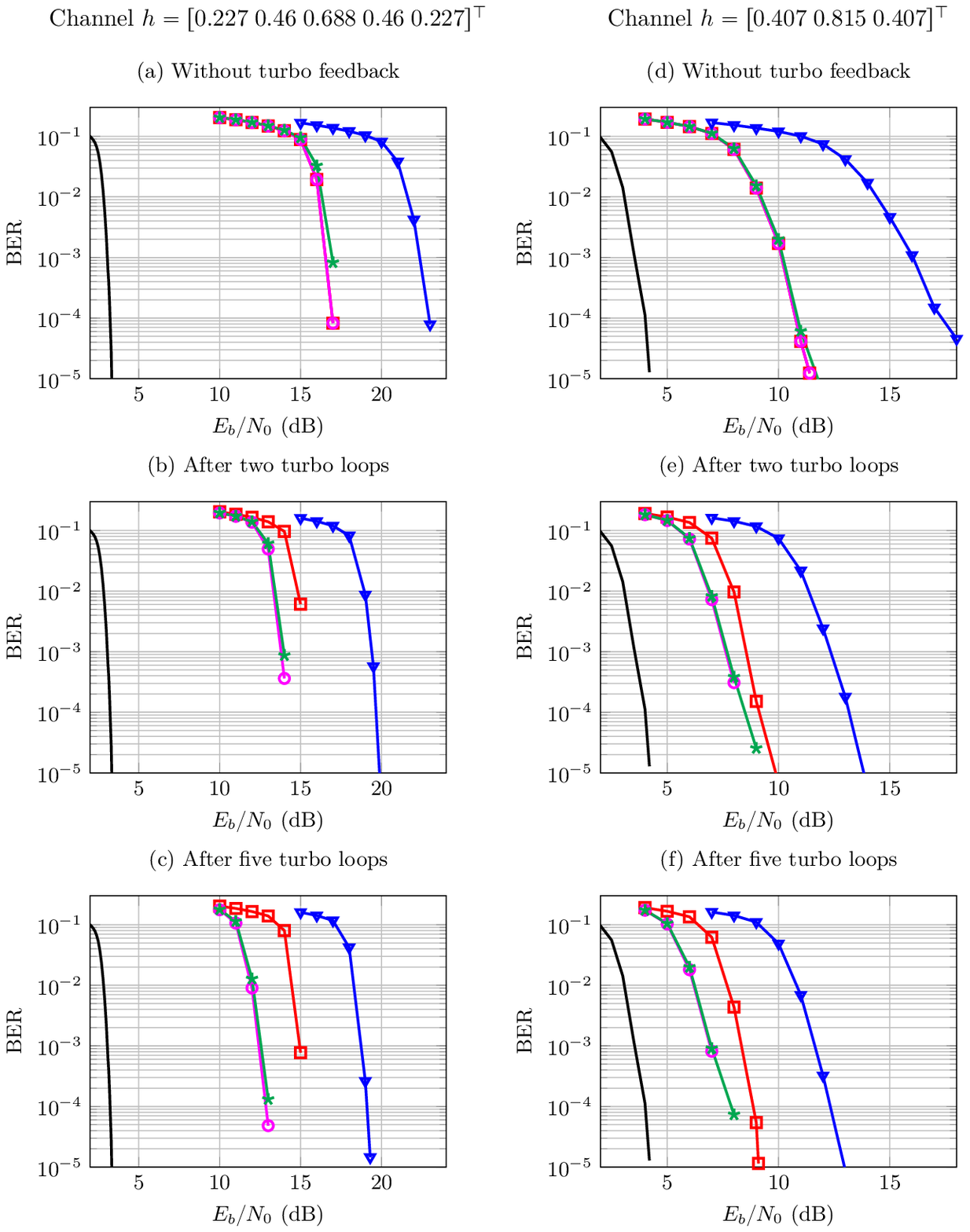}
%\scalebox{0.9}{\input{pics/BER8PSKeps1e-8Gray.tex}}
\caption{\small BER along $\EbNo$ for BEP \cite{Santos16} (\textcolor{red}{$\square$}), nuBEP (\textcolor{mycolor3}{$\circ$}), EP-F (\textcolor{mycolor}{$\star$}) and block-LMMSE (\textcolor{blue}{$\smalltriangledown$}) turbo equalizers, 8-PSK, codewords of $\t=4096$ (a)-(c) and $\t=1024$ (d)-(f) and two different channel responses. Black lines represent the AWGN bound. }\LABFIG{8PSK}
\end{figure*} 

In \FIG{8PSK} we simulate the same scenario \notaI{of} \FIG{BPSK}, but using an 8-PSK modulation rather than a BPSK. %, but averaging over 100 random channels and $1e^3$ random frames per channel realization. 
%We have increased the length of the encoded words to $\t=4098$ bits in \FIG{8PSK} in order to simulate exactly the same scenario than in  \cite{Singer01}. 
It can be observed that after increasing the order of the modulation, the EP-F approach presented in this work {performs identically as} its block {counterpart}, greatly improving the performance of the LMMSE algorithm before and after the turbo procedure. An improvement of the BER of the EP-F with respect to the one of the BEP approach in \cite{Santos16}, after turbo equalization, can also be observed.

In \FIG{1664QAM} we depict the BER performance after five turbo loops for channels $\vect{h}=[0.227\;0.46\;0.688\;0.46\;0.227]\trs$ in (a) and $\vect{h}=[0.407\;0.815\;0.407]\trs$ in (b) with different modulations. We use solid lines to represent a 64-QAM constellation and dashed lines for a 16-QAM. We sent codewords of length $\t=4096$ in both scenarios. It can be observed that the performance of the EP-F matches with the one of its block implementation proposed in this paper (nuBEP) when a 16-QAM is used. However, the EP-F approach slightly degrades with a 64-QAM, where the block nuBEP gets the most accurate performance. Note that the behavior of the \mbox{EP-F} could be improved by increasing the length of the filter, yielding the performance of its block implementation. We have a remarkable improvement of $3$-$5$ dB with respect to the BEP in \cite{Santos16} and of $7$-$13$ dB compared to the LMMSE algorithm.

\begin{figure*}[htb]
\centering
\includegraphics[width=7.0625in]{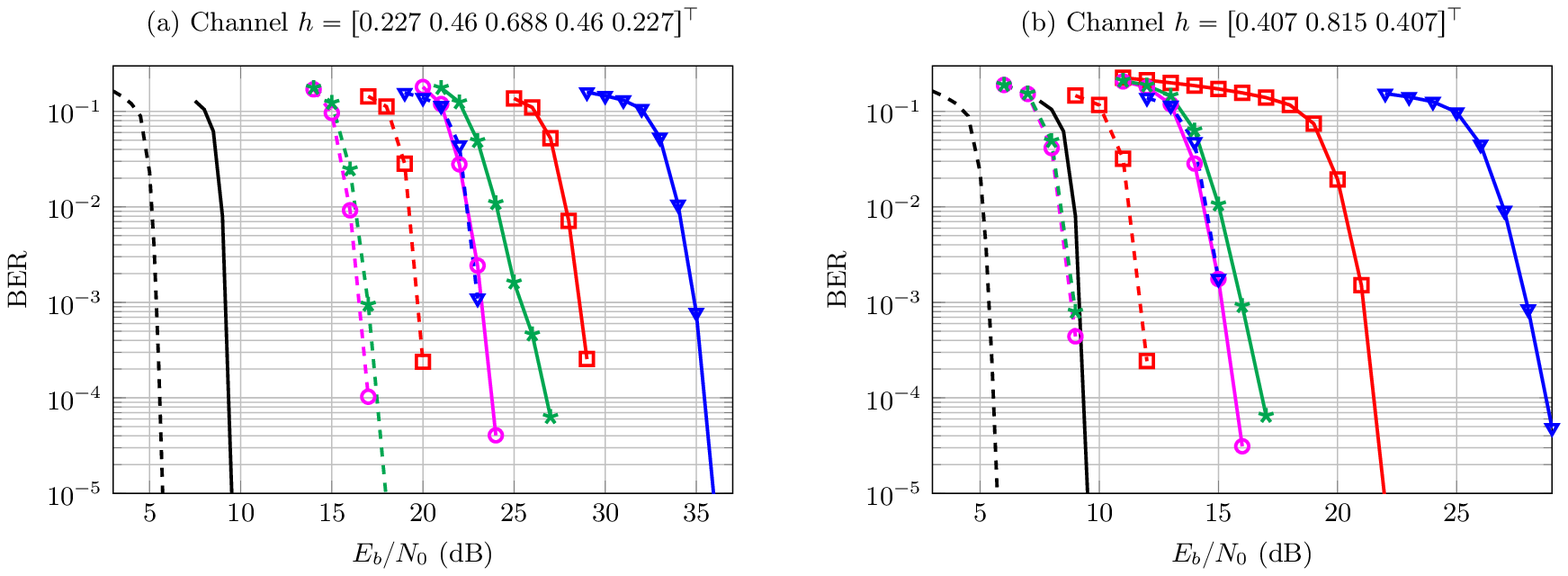}
%\scalebox{0.9}{\input{pics/BER16QAM64QAMGrayturbo1e-8.tex}}
\caption{\small BER along $\EbNo$ for BEP \cite{Santos16} (\textcolor{red}{$\square$}), nuBEP (\textcolor{mycolor3}{$\circ$}), EP-F (\textcolor{mycolor}{$\ast$}) and block-LMMSE (\textcolor{blue}{$\smalltriangledown$}) turbo equalizers after five turbo loops, 64-QAM (solid lines) and 16-QAM (dashed lines), codewords of $\t=4096$ and two different channel responses. Black lines represent the AWGN bound. }\LABFIG{1664QAM}
\end{figure*} 

For the sake of completeness, we include \FIG{8PSKDifV} to  show how the BER changes along the turbo iterations and different block lengths at $\EbNo=13$ dB for an 8-PSK and $\vect{h}=[0.227\;0.46\;0.688\;0.46\;0.227]\trs$. The nuBEP algorithm is represented in (a) and the filter approach EP-F in (b). It can be observed that BER is higher for shorter codes and it improves when the code length is increased, as expected. Also, the BER does not significantly  improve after the fifth turbo iteration. In the view of these results we stopped running our turbo equalizers after five turbo iterations in the experiments presented above. 

\begin{figure*}[htb]
\centering
\includegraphics[width=7.0625in]{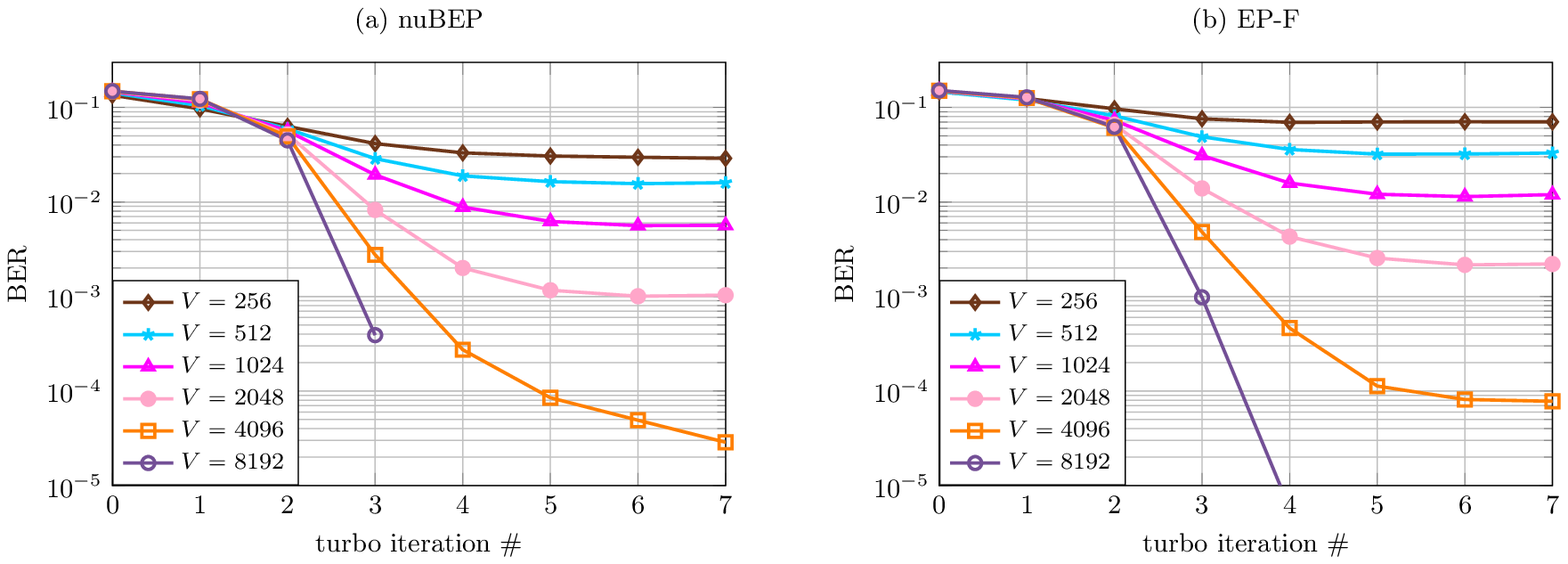}
%\scalebox{0.9}{\input{pics/BERDifV8PSKGray.tex}}
\caption{\small BER of nuBEP (a) and EP-F (b) turbo equalizers at $\EbNo=13$ dB for several turbo iterations and lengths of encoded words. 8-PSK modulation and the channel response $\vect{h}=[0.227\;0.46\;0.688\;0.46\;0.227]\trs$ were used. }\LABFIG{8PSKDifV}
\end{figure*}

\section{Conclusion}\LABSEC{conc}

In a previous work, we presented a novel equalizer based on expectation propagation (EP) \cite{Santos16}. This solution presents quite an improved performance compared to previous approaches in the literature, both for hard, soft and  turbo detection. The solution was presented as a block-wise solution and it was therefore denoted as block-EP (BEP). 
%Its major advantage is not having an exponential growth of computational complexity with the constellation size and the channel memory, where most of the equalizers are not available. 
The major advantage of the BEP lies in the fact that its computational complexity does not grow  exponentially with the constellation size and channel memory, as opposed to most equalizers, which are unfeasible for moderate values of these parameters. 
However, it exhibits a quadratic increase with the size of the transmitted word, $\t$. To avoid this problem, filter-type equalizers are usually preferred \cite{Berrou10}. For this reason, we proposed a smoothing EP (SEP) equalizer in \cite{Santos17}. However, the SEP has a computational complexity cubic in the channel length, $\ntaps$. Both BEP and SEP equalizers make use of a moderate feedback in the sense that an initial uniform discrete prior is assumed at the beginning of each execution of the EP algorithm, even after the turbo procedure has started. 
%However, both BEP and SEP equalizers assume uniform discrete prior from the decoder, even when the turbo procedure starts. 
In this paper, we first propose a design to include the non-uniform discrete nature of the priors from the decoder in the EP algorithm, which amounts to a stronger feedback, quite outperforming the previous BEP and SEP approaches. Second, we develop a reduced-complexity approach by proposing better values of the EP parameters. The resulting algorithm has been denoted as nuBEP, and it significantly outperforms the BEP reducing the computational complexity to less than the third part. Finally, we adapt the EP block equalizer to the filter-type form, emulating the Wiener MMSE filter-type \cite{Tuchler11}. Therefore, we mimic the structure of the filter-type MMSE equalizer. The EP is used to better approximate the posteriors of a windowed version of the inputs, shifted for every new output estimate. As a result, we present a novel solution dealing with $\winsize$ inputs at a time and with quadratic computational complexity in $\winsize$. This novel solution, the EP-F, despite the reduction in the computational complexity, exhibits a performance in terms of BER quite close to that of its block counterpart, the nuBEP. Furthermore, it remarkably improves the performance of the LMMSE turbo-equalizer, with same complexity order in terms of $\ntaps$ and $\t$. In the included experiments, for channels usually used as benchmarks in the literature, gains in the range $5$-$13$ dB are reported for $8$-PSK, $16$-QAM and $64$-QAM modulations. 

\notajj{One of the main benefits of this new proposal is to reduce the computational complexity, reducing it to be of quadratic order with the filter length. Other approaches, such as those solutions working on the frequency domain \cite{Karjalainen07}, could be investigated to achieve this goal. In this paper we face the equalization in single-input single-output channels, the application to MIMO channels with memory \cite{Xiao17} remains unexplored.  } 

%In the experimental results included, the EP-F always improves the performance of the turbo LMMSE approach. In addition, EP filter and its block implementation share quite close performance even for high-order constellations and outperform the LMMSE algorithm even before the turbo equalization. We also show that the block EP implementation, nuBEP, proposed in this paper, quite outperforms the BEP approach in \cite{Santos16}. 

% if have a single appendix:
%\appendix[Proof of the Zonklar Equations]
% or
%\appendix  % for no appendix heading
% do not use \section anymore after \appendix, only \section*
% is possibly needed

% use appendices with more than one appendix
% then use \section to start each appendix
% you must declare a \section before using any
% \subsection or using \label (\appendices by itself
% starts a section numbered zero.)
%

%
\appendices
\section{Proof of \EQ{posmeanep} and \EQ{posvarnep}}\LABAPEN{ap1}
%Appendix one text goes here.

In \cite{Santos16}, the posterior distribution used for BEP is
\begin{equation}
q^{[\ell]}(\vect{\beforechannel})\sim\cgauss{\vect{\beforechannel}}{\boldsymbol{\mu}^{[\ell]}}{\boldsymbol{\Sigma}^{[\ell]}}
\end{equation}
where
\begin{align}
\boldsymbol{\mu}^{[\ell]} & = \boldsymbol{\Sigma}^{[\ell]}(\sigma_\noise^{-2}\matr{H}\her\vect{\afterchannel}+\mbox{diag}(\notaI{\boldsymbol{\eta}}^{[\ell]})\inv\vect{m}^{[\ell]}), \LABEQ{dem1}\\
\boldsymbol{\Sigma}^{[\ell]} &= \left(\sigma_\noise^{-2}\matr{H}\her\matr{H}+\mbox{diag}(\notaI{\boldsymbol{\eta}}^{[\ell]})\inv\right)\inv. \LABEQ{dem2}
\end{align}
By a direct application of the Woodbury identity, equation \EQ{dem2} can be rewritten as
\begin{equation}
\boldsymbol{\Sigma}^{[\ell]} = \mbox{diag}(\notaI{\boldsymbol{\eta}}^{[\ell]})-\mbox{diag}(\notaI{\boldsymbol{\eta}}^{[\ell]})\matr{H}\her \matr{C}\inv\matr{H}\mbox{diag}(\notaI{\boldsymbol{\eta}}^{[\ell]}) \LABEQ{dem3}
\end{equation}
where
\begin{equation}
\matr{C}=\matr{H}\mbox{diag}(\notaI{\boldsymbol{\eta}}^{[\ell]})\matr{H}\her + \sigma_\noise^2\matr{I}.
\end{equation}
The $k$-th diagonal element of \EQ{dem3} yields \EQ{posvarnep}. 
Regarding \EQ{dem1}, it can be divided into two terms
\begin{equation}
\boldsymbol{\mu}^{[\ell]} = \underbrace{\boldsymbol{\Sigma}^{[\ell]}\sigma_\noise^{-2}\matr{H}\her\vect{\afterchannel}}_{T_1}+\underbrace{\boldsymbol{\Sigma}^{[\ell]}\mbox{diag}(\notaI{\boldsymbol{\eta}}^{[\ell]})\inv\vect{m}^{[\ell]}}_{T_2}.\LABEQ{dem4}
\end{equation}
We apply the following identity \cite{Bishop06}, 
\begin{equation}
(\matr{A}\inv+\matr{B}\her\matr{D}\inv\matr{B})\inv\matr{B}\her\matr{D}\inv=\matr{A}\matr{B}\her (\matr{B}\matr{A}\matr{B}\her+\matr{D})\inv
\end{equation}
to the first term, $T_1$, in \EQ{dem4}, yielding 
\begin{equation}
T_1=\mbox{diag}(\notaI{\boldsymbol{\eta}}^{[\ell]})\matr{H}\her\matr{C}\inv\vect{\afterchannel}. \LABEQ{dem5}
\end{equation}
Now, we replace \EQ{dem3} into the second term, $T_2$, in \EQ{dem4}, obtaining
\begin{equation}
T_2= \vect{m}^{[\ell]}-\mbox{diag}(\notaI{\boldsymbol{\eta}}^{[\ell]})\matr{H}\her\matr{C}\inv\matr{H}\vect{m}^{[\ell]}. \LABEQ{dem6}
\end{equation}
By replacing \EQ{dem5} and \EQ{dem6} into \EQ{dem4}, we finally get
\begin{equation}
\boldsymbol{\mu}^{[\ell]} = \vect{m}^{[\ell]}+\mbox{diag}(\notaI{\boldsymbol{\eta}}^{[\ell]})\matr{H}\her\matr{C}\inv(\vect{\afterchannel}-\matr{H}\vect{m}^{[\ell]}),
\end{equation}
whose $\iter$-th element is given by  \EQ{posmeanep}. 

\section{Proof of \EQ{mean_extF} and \EQ{var_extF}}\LABAPEN{ap2}

In \cite{Koetter02}, the extrinsic distribution of the estimated symbol is computed as
\begin{equation}\LABEQ{dem7}
q_E(\widehat{\beforechannel}_\iter|\beforechannel_\iter)\sim\cgauss{\widehat{\beforechannel}_\iter}{\beforechannel_\iter\vecti{\vcoef}{\iter}\her\vecti{h}{\winsize}}{\sigma_\iter^2}
\end{equation}
where
\begin{align}
\widehat{\beforechannel}_\iter &= \vecti{\vcoef}{\iter}\her(\vecti{\afterchannel}{\iter} - \vecti{H}{\winsize}\vecti{m}{\iter}+m_\iter\vecti{h}{\winsize}),\\
\sigma_\iter^2&= \vecti{\vcoef}{\iter}\her\vecti{h}{\winsize}\energy(1-\vecti{h}{\winsize}\her\vecti{\vcoef}{\iter}),
\end{align}
$\vecti{\vcoef}{\iter}$ is given by \EQ{vcoef} and $\vecti{h}{\winsize}$ is the ($\winsize_2+\ntaps$)-th column of $\vecti{H}{\winsize}$ defined in \EQ{Hk}. Note that we generalized the expressions in \cite{Koetter02} to consider a symbol energy of $\energy$. If we set $\energy=1$, we obtain  exactly the formulation in  \cite{Koetter02}. 
Instead of the extrinsic distribution of the estimated symbol, we use in our formulation the extrinsic distribution of the true symbol, which can be computed from \EQ{dem7} as
\begin{equation}
q_E({\beforechannel}_\iter|\widehat{\beforechannel}_\iter)\sim\cgauss{\beforechannel_\iter}{z_\iter}{\notaI{v_\iter^2}}
\end{equation}
where
\begin{align}
z_\iter &= \frac{\widehat{\beforechannel}_\iter}{\vecti{\vcoef}{\iter}\her\vecti{h}{\winsize}},\\
\notaI{v_\iter^2}&= \frac{\sigma_\iter^2}{(\vecti{\vcoef}{\iter}\her\vecti{h}{\winsize})^2},
\end{align}
yielding the formulation in \EQ{mean_extF} and \EQ{var_extF}. 

% you can choose not to have a title for an appendix
% if you want by leaving the argument blank
%\section{}
%Appendix two text goes here.

% use section* for acknowledgment
%\section*{Acknowledgment}
%
%
%The authors would like to thank...

% Can use something like this to put references on a page
% by themselves when using endfloat and the captionsoff option.
\ifCLASSOPTIONcaptionsoff
  \newpage
\fi

% trigger a \newpage just before the given reference
% number - used to balance the columns on the last page
% adjust value as needed - may need to be readjusted if
% the document is modified later
%\IEEEtriggeratref{8}
% The "triggered" command can be changed if desired:
%\IEEEtriggercmd{\enlargethispage{-5in}}

% references section

% can use a bibliography generated by BibTeX as a .bbl file
% BibTeX documentation can be easily obtained at:
% http://mirror.ctan.org/biblio/bibtex/contrib/doc/
% The IEEEtran BibTeX style support page is at:
% http://www.michaelshell.org/tex/ieeetran/bibtex/
%\bibliographystyle{IEEEtran}
% argument is your BibTeX string definitions and bibliography database(s)
%\bibliography{IEEEabrv,../bib/paper}
%
% <OR> manually copy in the resultant .bbl file
% set second argument of \begin to the number of references
% (used to reserve space for the reference number labels box)
%\begin{thebibliography}{1}
%
%\bibitem{IEEEhowto:kopka}
%H.~Kopka and P.~W. Daly, \emph{A Guide to \LaTeX}, 3rd~ed.\hskip 1em plus
%  0.5em minus 0.4em\relax Harlow, England: Addison-Wesley, 1999.
%
%\end{thebibliography}

\bibliographystyle{IEEEtran}
\bibliography{allBib}

\end{document}